\RequirePackage{fix-cm}
\documentclass{svjour3}                     
\smartqed  
\usepackage{graphicx,url}
\usepackage{epstopdf}

\usepackage{epsfig,textcomp}
\usepackage{amssymb, amsfonts, amsmath}
\usepackage{float}
\usepackage{color, multirow, enumitem}
\begin{document}

\title{Prediction of air pollutants PM$_{10}$ by   ARBX(1) processes \thanks{This work was supported in part by project MTM2015--71839--P (co-funded by Feder funds), of the DGI, MINECO, Spain.}
}

\titlerunning{Functional prediction of air pollutants PM$_{10}$}        

\author{J. \'Alvarez-Li\'ebana         \and
        M. D. Ruiz-Medina 
       }


\institute{J. \'Alvarez-Li\'ebana \at
              Department of Statistics, O.R. and Didactics of Mathematics, University of Oviedo \\
              \email{alvarezljavier@uniovi.es}           
           \and
           M. D. Ruiz-Medina \at
             Department of Statistics and O.R., University of Granada  \email{mruiz@ugr.es}
}

\date{        \vspace{-0.3cm}
 Received: date / Accepted: date}

\maketitle

\begin{abstract}
This work adopts a Banach-valued time series framework for   estimation and prediction, from temporal correlated functional data, in presence of exogenous variables.    The consistency of the proposed  functional predictor  is illustrated in the simulation study undertaken.   Air pollutants PM$_{10}$ 
curve forecasting, in the Haute--Normandie region (France), is addressed, by implementation of  the functional time series approach presented.

\keywords{Air pollutants forecasting \and Banach spaces \and   functional time series \and meteorological variables \and strong consistency}
\end{abstract}


\section{Introduction}
\label{sec:1} 

\textcolor{blue}{In the literature,} several approaches have been adopted in the analysis of pollution data (see, e.g., 
\cite{Pang09}, for a comparative study). In \cite{Zhangetal18}, the singular value decomposition is applied to identify  spatial air pollution index (API) patterns, in relation  to meteorological conditions in China.  A novel hybrid model\textcolor{blue}{,} combining Multilayer perceptron model and Principal Component
Analysis (PCA)\textcolor{blue}{,} is introduced in \cite{He15}, to improve the air quality prediction accuracy  
 in urban areas. Factor analysis and 
 Box--Jenkins methodology are considered  in \cite{Gocheva14}, to examine concentrations of primary air pollutants such as NO, NO\textcolor{blue}{$_{2}$} , NO$_{x}$ , PM$_{10},$ SO\textcolor{blue}{$_{2}$} and ground level O\textcolor{blue}{$_{3}$} in the town of Blagoevgrad, Bulgaria.
 Since PM$_{10}$ are inhalable atmospheric  particles, \textcolor{blue}{their forecast} has became crucial aimed at adopting efficient  public transport policies. In the recent literature, one can find several modelling approaches for 
 PM$_{10}$ forecasting. Among the most common statistical techniques applied, we mention  multiple regression, non--linear state space modelling and  artificial neural networks  (see, e.g., \textcolor{blue}{\cite{GrivasChaloulakou06,Paschalidouetal11,Slinietal06,Stadloberetal08,Zolghadrietal06}}). 
Functional Data Analysis (FDA) techniques also play a crucial role in air quality forecasting (see \cite{Febreroetal08,Fernandezetal05,Ignaccoloetal14}, among others).  
Related meteorological variables can also be functional predicted from a functional time series framework (see, e.g., 
\cite{Besseetal00,RuizEspejo12,RuizMedinaetal14}).

Computational advances have made possible the implementation of flexible  models for random elements in function spaces.
FDA techniques have emerged  in the local  analysis of high--dimensional data, which are functional in nature (see  the monographs 
\cite{GoiaVieu16,HorvathKokoszka12,HsingEubank15}, and the references therein). 
Parametric   functional   linear time series techniques are fast and computational   low-cost. In contrast with the more flexible nonparametric functional statistical approach (see, e.g.,\cite{FerratyVieu06}), 
 the so--called curse of dimensionality problem (see \cite{Geenens11,Vieu18})  is mitigated. The  semi-parametric framework  also provides a partial solution to this problem (see, e.g.,  \cite{AneirosVieu08,GoiaVieu15}, in the 
 semi--functional regression context). 

 From a theoretical point of view, parametric   functional   linear time series techniques have been widely studied in the last few decades. In particular, in the autoregressive Hilbertian process framework, the asymptotic properties of  componentwise  estimators of the autocorrelation operator, and their associated plug-in predictors 
have been derived in \cite{Bosq00,Mas04,Mas07}, among others. 
Recently, in  \cite{Alvarezetal17} and \cite{RuizAl19}, \textcolor{blue}{alternative} operator norms for  consistency have been investigated.  See also \cite{Alvarezetal16}, for the case of 
Ornstein--Uhlenbeck process in Hilbert and Banach spaces. 

Linear parametric  approaches have also been adopted in a Banach-valued time series framework.  This literature has mainly been focused on  the spaces of continuous functions $\mathcal{C}([0,1])$  with the supremum norn (see \cite{BuenoKlepsch18,DehlingSharipov05,LabbasMourid02,Parvardehetal17}, among others), and  on the Skorokhod space  of right--continuous      functions on $[0,1],$ having limit to the left at each $t\in [0,1],$ equipped with the Skorokhod topology, usually denoted as $\mathcal{J}_{1}$--topology (see, e.g., 
\cite{BlankeBosq16,Hajj11}).
 The lack of an inner product structure is supplied in \cite{RuizAlvarez18_JMVA}, by considering suitable embeddings into related Hilbert spaces.  \textcolor{blue}{In the Banach context, authors mostly focused on finding a finer scaler of norms for measuring the  local regularity.} Specifically, this paper proves strong consistency of a componentwise estimator of the autocorrelation operator and its associated plug--in predictor, in an abstract Banach--valued time series framework\textcolor{blue}{, under the opposite motivation: measuring the local singularity in an accurately way.}

 A first attempt for  the inclusion of exogenous information in the  functional  time series framework    can be found  in  \textcolor{blue}{\cite{DamonGuillas02}}, where  the so-called ARHX(1) processes  (Hilbert-valued autoregressive processes of order one with exogenous variables) are introduced. 
  Enhancements were subsequently proposed by  \textcolor{blue}{\cite{DamonGuillas05,MarionPumo04}}. First order conditional autoregressive Hilbertian processes  were introduced in \cite{Guillas02}.   
 The present paper extends the time series 
 framework in \cite{RuizAlvarez18_JMVA} to the case of first-order Banach-valued autoregressive processes with exogenous variables (ARBX(1) processes). Functional parameter estimation and plug--in prediction can be addressed in our ARBX(1) context,  from the multivariate infinite--dimensional formulation of the results 
 in \cite{RuizAlvarez18_JMVA}.
 Specifically, a matrix--operator--based formulation of the ARB(1) process (Banach--valued autoregressive process of order one) state equation  is considered. The required Hilbert space embeddings, and sufficient conditions for the strong--consistency of the autocorrelation operator estimator (reflecting temporal correlations between endogenous and exogenous variables), and the associated plug--in functional predictor\textcolor{blue}{,} are then  obtained in a direct way.   We refer to the reader to \cite{Triebel83}, where several examples of the  Banach space context  introduced in \cite{RuizAlvarez18_JMVA} can be found.

The outline of the paper is as follows. The ARBX(1) based estimation and prediction methodologies presented are described in Section \ref{sec:2}. \textcolor{blue}{Some basic concepts about nuclear and Besov spaces are introduced in Section \ref{sec:3}, as well as a discussion about how assumptions are verified in the scenarios adopted.} A simulation study is undertaken, to illustrate the consistency of ARBX(1) predictors, in Section \ref{sec:4}.  PM$_{10}$ short-term forecasting, based on the introduced ARBX(1) framework,  is addressed in Section \ref{sec:5}. Final comments, \textcolor{blue}{and discussion about how the extension to an spatiotemporal analysis,} are provided in Section \ref{sec:6}. \textcolor{red}{Some extra Figures and Tables can be found in the Appendix. The main theoretical proofs required for the development of the results, as well as more detailed definitions of Besov spaces, can be checked in the Supplementary Material provided.}


\section{ARBX(1) estimation and prediction}
\label{sec:2}

In the following, the functional random variables and stochastic processes introduced below are defined on the basic probability space
$\left( \Omega, \mathcal{A}, \mathcal{P} \right).$ Let $\left(B, \left\| \cdot \right\|_B \right)$ be a real separable Banach space with associated norm $\left\| \cdot \right\|_B.$  
 Consider   $X = \left\lbrace X_n, \ n \in \mathbb{Z} \right\rbrace$ to be a zero-mean ARB(1) process, with $P[X_{n}\in B]=1,$ $n\in \mathbb{Z},$  satisfying the following state equation (see, e.g.,  \cite{Bosq00}):

\begin{equation}
X_n  = \rho \left( X_{n-1} \right)  + \varepsilon_n , \quad  n \in \mathbb{Z},\label{state_equation_Banach}
\end{equation}
\noindent  where $\rho $ is the autocorrelation operator, which is assumed to be a bounded linear operator on $B,$ that is, $\rho \in \mathcal{L}(B),$  with $\left(\mathcal{L}(B), \left\| \cdot \right\|_{\mathcal{L}(B)} \right)$ denoting the Banach space of continuous operators with the supremum norm.  Here, $\varepsilon = \left\lbrace \varepsilon_n, \ n \in \mathbb{Z} \right\rbrace$  represents the  innovation process, which is assumed to be a $B$-valued strong white noise, and uncorrelated with the random initial condition. In particular, $\sigma_{\varepsilon}^{2} = {\rm E}  \textcolor{blue}{\left[ \left\| \varepsilon_n \right\|_{B}^{2} \right]} < \infty,$ $n\in \mathbb{Z}.$ From \cite[Theorem 6.1]{Bosq00}, if there exists  $j_0 \geq 1$ such that $\left\| \rho^{j} \right\|_{\mathcal{L}\left(B \right)} < 1,$ for every $j \geq j_0,$
then, equation (\ref{state_equation_Banach}) admits a unique stationary solution  $X_n =  \sum_{j=0}^{\infty} \rho^j \left( \varepsilon_{n-j} \right)$, with $\sigma_{X}^{2} = {\rm E}  \textcolor{blue}{\left[ \left\| X_n \right\|_{B}^{2} \right]} < \infty,$ $n \in \mathbb{Z}.$ 

In this paper, exogenous information is incorporated to equation (\ref{state_equation_Banach}) in an additive way. Thus, the state equation of an ARBX(1) process will be established as follows:
 \begin{equation}
X_n  = \rho \left( X_{n-1} \right)  + \displaystyle \sum_{i=1}^{b}a_i\left( Z_{n,i} \right)   + \varepsilon_n , \quad n \in \mathbb{Z}, \label{state_equation_ARBX}
\end{equation}
\noindent where $\left\lbrace a_i, \ i=1,\ldots,b \right\rbrace$ are bounded linear operators on $B.$ The exogenous functional random variables $Z_i=\{ Z_{n,i},\ n\in \mathbb{Z}\},$ $i=1,\ldots, b,$  are assumed to satisfy the following ARB(1)  equation, for $i=1,\ldots, b,$ 
\begin{equation}
Z_{n,i}  =  u_{i} \left( Z_{n-1,i} \right)  + \eta_{n,i},  \quad u_{i} \in \mathcal{L}(B), \quad n \in \mathbb{Z}.  \label{eq_11}
\end{equation}
\noindent  For $i=1,\dots,b,$ $\eta_i = \left\lbrace \eta_{n,i} \ n \in \mathbb{Z} \right\rbrace$ is a $B$-valued strong white noise. Particularly,  $\sigma_{\eta_{i}}^{2} = {\rm E}\left[ \left\| \eta_{n,i} \right\|_{B}^{2} \right] < \infty,$ $n\in \mathbb{Z},$ $i=1,\dots,b.$ 
Here,   
 $P[Z_{n,i}\in B]=1, \quad {\rm E} \left[ Z_{n,i} \right] = 0_{B},$   $n \in \mathbb{Z},$   for $i=1,\ldots,b.$ The symbol $0_{B}$  means the zero element  \textcolor{blue}{(null function) in  $B.$} Equations (\ref{state_equation_ARBX})--(\ref{eq_11}) can be rewritten as 
   (see \cite{DamonGuillas02}),  
   
  \begin{eqnarray}
\overline{X}_n &=& \overline{\rho} \left( \overline{X}_{n-1} \right) + \overline{\varepsilon}_n, \quad \overline{\rho} \in \mathcal{L}(\overline{B}), \quad P[\overline{X}_n\in \overline{B}]=P[\overline{\varepsilon}_n \in \overline{B}]=1, 
\label{maarb1eq}
\end{eqnarray}
\noindent where $\overline{B}=B^{b+1}$ \textcolor{red}{is also a real separable Banach space (see Lemma 1 and Proposition 1 in the Supplementary Material provided)}, and 
\begin{equation}
\overline{X}_n = \begin{pmatrix} X_n \\ Z_{n+1,1} \\ Z_{n+1,2} \\ \vdots \\  Z_{n+1,b} \end{pmatrix}, \quad \overline{\varepsilon}_{n}= \begin{pmatrix} \varepsilon_n \\ \eta_{n,1} \\ \eta_{n,2} \\ \vdots\\  \eta_{n,b} \end{pmatrix},  \quad \overline{\rho} = \begin{pmatrix} \rho & a_1 & \textcolor{blue}{a_2} & \textcolor{blue}{a_3}  & \cdots & \textcolor{blue}{a_{b-1}} & a_b \\ \mathbf{0}_B & u_{1} & \mathbf{0}_B & \textcolor{blue}{\mathbf{0}_B } & \cdots   & \textcolor{blue}{\mathbf{0}_{B}} & \mathbf{0}_{B} \\  
\mathbf{0}_B & \mathbf{0}_B & u_{2} & \mathbf{0}_B & \textcolor{blue}{\cdots} & \textcolor{blue}{\mathbf{0}_B}  & \textcolor{blue}{\mathbf{0}_B} \\ 
\vdots & \vdots & \textcolor{blue}{\vdots} & \textcolor{blue}{\vdots} &\ddots & \textcolor{blue}{ \vdots} &
\textcolor{blue}{\vdots} \\ \textcolor{blue}{\mathbf{0}_B} & \textcolor{blue}{\mathbf{0}_{B}}  &\textcolor{blue}{\mathbf{0}_{B}} & \textcolor{blue}{\mathbf{0}_{B}} & \textcolor{blue}{\cdots}  & \textcolor{blue}{\mathbf{0}_{B}} & \textcolor{blue}{\mathbf{0}_{B}} \\ \textcolor{blue}{\mathbf{0}_B} & \textcolor{blue}{\mathbf{0}_{B}}  &\textcolor{blue}{\mathbf{0}_{B}} & \textcolor{blue}{\mathbf{0}_{B}} & \textcolor{blue}{\cdots}  & \textcolor{blue}{u_{b-1}} & \textcolor{blue}{\mathbf{0}_{B}} \\ \mathbf{0}_B & \mathbf{0}_{B}  &\textcolor{blue}{\mathbf{0}_{B}} & \mathbf{0}_{B} & \cdots  & \textcolor{blue}{\mathbf{0}_{B}} & u_{b} \end{pmatrix}. 
   \label{eq_13}\end{equation}
 \noindent Here, $\mathbf{0}_B$ represents the null operator on $B.$
In equation (\ref{maarb1eq}), $\mathcal{L}(\overline{B})$ denotes the space of bounded linear operators on $\overline{B},$  endowed with the norm
\begin{equation}
\left\| \overline{y} \right\|_{\overline{B}}  = \displaystyle \sup_{n \geq 1} \left| \overline{F}_n (\overline{y})
\right|=\sup_{n\geq 1}\sup_{i \in \{1,\dots, 
b+1\}}\quad \left|F_{ni}(y_{i})\right|,\quad \overline{F}_{n}=\left(F_{n1},\dots,F_{n(b+1)}\right),\label{veBnorm}\end{equation}

\noindent for every $\overline{y}=(y_{1},\dots,y_{b+1})\in \overline{B}=B^{b+1}.$ Here, for $i=1,\dots,b,$ 
$\{ F_{ni},\  n \geq 1 \} \subset B^{\star}$ is a sequence of bounded linear functionals on $B$ satisfying
\begin{equation}
F_{ni} \left( x_{ni} \right) = \left\| x_{ni} \right\|_B, \quad \left\| F_{ni} \right\| =  1,\quad n\geq 1,\label{eq_16}
\end{equation}
\noindent with $\{ x_{ni},\  n \geq 1 \} \subset B$ being a dense sequence in $B$ (see Lemma 2.1
in \cite{Kuelbs70} for more details). For simplification purposes, we consider a common dense system in $B,$ i.e.,  $x_{n}=x_{ni},$  and $F_{ni}=F_{n},$ for $i=1,\dots,b,$ and $n\geq 1.$ We will assume the conditions ensuring the existence of a unique stationary solution to  equation (\ref{maarb1eq}). That is, assume that  there exists a $j_{0}$ such that  $\|\overline{\rho}^{j}\|_{\mathcal{L}(\overline{B})}<1,$ for all $j\geq j_{0}.$  
The following 
 componentwise estimator of the autocorrelation operator  $\overline{\rho}$ in (\ref{maarb1eq}),       based on a functional sample of size $n,$ is then  formulated:
 \begin{equation}
\widetilde{\overline{\rho}}_{k_n} (\overline{x}) = \left( \widetilde{\Pi}^{k_n} \overline{D}_n \overline{C}_{n}^{-1} \widetilde{\Pi}^{k_n} \right) (\overline{x}) =  \left( \displaystyle \sum_{j=1}^{k_n} \frac{1}{C_{n,j}} \langle \overline{x}, \overline{\phi}_{n,j} \rangle_{\widetilde{\overline{H}}}\widetilde{\Pi}^{k_n} \overline{D}_n(\overline{\phi}_{n,j}) \right), \label{estimator}
\end{equation}

 \noindent where, for $j\geq 1,$  $$\langle \overline{x}, \overline{\phi}_{n,j} \rangle_{\widetilde{\overline{H}}}=\left\langle x, \phi_{n,j1}\right\rangle_{\widetilde{H}}+
 \sum_{i=1}^{b}\left\langle x_{i}, \phi_{n,j(i+1)}\right\rangle_{\widetilde{H}},\quad \forall \overline{x}=(x,x_{1},\cdots, x_{b}) \in \overline{B},$$ \noindent with $\{\overline{\phi}_{n,j}=(\phi_{n,j1},\cdots,\phi_{n,j(b+1)}),\ j\geq 1\}$ being the orthonormal \textcolor{blue}{eigenfunctions} system associated with $$\overline{C}_{n}=\frac{1}{n}\sum_{i=1}^{n}\overline{X}_{i}\otimes \overline{X}_{i},$$\noindent   the empirical autocovariance operator of the extended version to $\widetilde{\overline{H}}=\widetilde{H}^{b+1},$ of  $\overline{X}=\{\overline{X}_{n},\  n\in \mathbb{Z}\}.$
 Here, the Hilbert space
     $\widetilde{H}$ is   defined   in Lemma 2.1
in \cite{Kuelbs70}, as the continuous extension of  the separable 
real--valued Banach space $B$ (see also Lemma 1
in \cite{RuizAlvarez18_JMVA}). In particular, its inner product is given by $\left\langle f,g\right\rangle_{\widetilde{H}}=\sum_{n=1}^{\infty}t_{n}F_{n}(f)F_{n}(g),$ for $f,g\in \widetilde{H},$ with $\sum_{n=1}^{\infty }t_{n}=1,$ $t_{n}>0,$ $n\geq1.$  Note that  $\widetilde{H}$ has weaker topology than $B,$ allowing  the continuous  inclusion  $B\hookrightarrow \widetilde{H},$ and hence, 
$\overline{B}\hookrightarrow \widetilde{\overline{H}},$ holds. In (\ref{estimator}),
for each functional sample \textcolor{blue}{of} size  $n,$ we have denoted
\begin{eqnarray}\widetilde{\Pi}^{k_n} 
\left(\overline{x}\right)&=&\displaystyle \sum_{j=1}^{k_n} \langle \overline{x}, \overline{\phi}_{n,j} \rangle_{\widetilde{\overline{H}}} \overline{\phi}_{n,j},\quad \forall \overline{x}\in \overline{B}=B^{b+1}\subset \widetilde{\overline{H}}=\widetilde{H}^{b+1}.
\end{eqnarray}
\noindent Denote also by $\{C_{n,j},\ j\geq 1\},$ with $C_{n,1}\geq \dots\geq C_{n,n}\geq 0=C_{n,n+1}= C_{n,n+2}=\dots,$  the  eigenvalues of $\overline{C}_{n}$ respective  associated with the empirical \textcolor{blue}{eigenfunctions}  $\{\overline{\phi}_{n,j},\ j\geq 1\}.$ The operator  $\overline{D}_{n}=
\frac{1}{n-1}\sum_{i=1}^{n-1}\overline{X}_{i}\otimes\overline{X}_{i+1}$  denotes  the empirical 
cross--covariance operator of the  extended version of $\overline{X}_{n}$ to $\widetilde{\overline{H}}.$  

In \cite{RuizAlvarez18_JMVA}, sufficient conditions are derived to ensure the strong consistency in the space $\mathcal{L}(\overline{B})$ (i.e., with respect to the supremum norm in $\mathcal{L}(\overline{B})$) of the  componentwise functional parameter estimator (\ref{estimator}) of $\overline{\rho},$ formulated in the weaker topology of $\widetilde{\overline{H}},$ where countable  orthogonal systems, like $\{\overline{\phi}_{n,j},\ j\geq 1\},$ can be considered \textcolor{blue}{from its separability}.  Specifically\textcolor{blue}{,} the following conditions are assumed in  \cite{RuizAlvarez18_JMVA}:

\begin{itemize}[label=$\bullet$]

\item \textbf{Assumption A1.} $\| \overline{X}_{0}\|_{\overline{B}}$ is almost surely bounded. The  eigenspaces associated with the eigenvalues of  $\overline{C}=E\left[\overline{X}_{n}\otimes \overline{X}_{n}\right]$ are one--dimensional.

\item  
\textbf{Assumption A2}. Let $k_{n}$ be such that
$ C_{n,k_{n}}>0$ a.s., and both $k_{n}\to \infty$ and ${k_{n}}/{n}\to 0$ as $n\to \infty$. Here, $C_{n,k_{n}}$ denotes the $k_{n}$--th eigenvalue of $\overline{C}_{n}.$
\item 
\textbf{Assumption A3.}  As $k \to \infty$,

\begin{small}
$$
\sup_{\overline{x}\in \overline{B}, \ \|\overline{x}\|_{\overline{B}}\leq 1}\left\|\overline{\rho}(\overline{x})-\sum_{j=1}^{k}\langle\overline{\rho}(\overline{x}),\overline{\phi}_{j}\rangle_{\widetilde{\overline{H}}}\overline{\phi}_{j}\right\|_{\overline{B}}\to 0,\quad \overline{\phi}_{j}=(\phi_{j1},\dots,\phi_{j(b+1)}),\ j\geq 1,$$
\end{small}
\noindent where $\overline{C}(\overline{\phi}_{j})=C_{j}\overline{\phi}_{j},$ $j\geq 1,$ in
$\widetilde{\overline{H}}.$
\item \textbf{Assumption A4.} Denote by $\mathcal{H}(\overline{X})$ the reproducing kernel Hilbert space generated by $\overline{C}.$
The  inclusion of $\mathcal{H}(\overline{X})$ into  $\widetilde{\overline{H}}^{\star}$ is continuous, i.e.,
$\mathcal{H}(\overline{X}) \hookrightarrow \widetilde{\overline{H}}^{\star},$ is a continuous mapping,  where $\widetilde{\overline{H}}^{\star}$ denotes the dual Hilbert space of $\widetilde{\overline{H}}.$
\item \textbf{Assumption A5.}
The  embedding $i_{\mathcal{H}(\overline{X}),\overline{H}}: \mathcal{H}(\overline{X})\hookrightarrow \overline{H}$ is   Hilbert--Schmidt. Here, $\overline{H}=H^{b+1},$ with $H$ being  a real--valued  separable Hilbert space   such that $\widetilde{\overline{H}}^{\star}\hookrightarrow \overline{H}\hookrightarrow \widetilde{\overline{H}}$ conforms a Rigged Hilbert space structure \textcolor{blue}{(also known as Gelfand triple)}.
\end{itemize}
Under \textbf{Assumptions A1--A5}, and the conditions assumed in \textcolor{blue}{\cite[Lemmas 8--9]{RuizAlvarez18_JMVA}} (see also the conditions imposed in  \textcolor{blue}{\cite[Theorem 1]{RuizAlvarez18_JMVA})}, 
  $$\left\| \overline{\rho} - \widetilde{\overline{\rho}}_{k_n} \right\|_{\mathcal{L} \left(\overline{B} \right)} \to^{a.s.} 0, \quad \left\| \left(\overline{\rho} - \widetilde{\overline{\rho}}_{k_n} \right) (\overline{X}_n) \right\|_{\overline{B}} \to^{a.s.}0,$$
\noindent where $\to_{a.s.}$  means the almost surely convergence.

\section{\textcolor{red}{A particular example: the scale of Besov spaces}}
\label{sec:3}

\textcolor{red}{This section is intended to provide the reader the main details about a particular Banach-valued framework in which the assumptions above displayed can be easily verified, in the context of nuclear spaces: the continuous scale of Besov spaces. The general model and tuning parameters adopted in the simulation study (see Section  \ref{sec:4} below), and the scenarios therein simulated, will be also discussed.}

\subsection{\textcolor{red}{Besov spaces and continuous embeddings}}

\textcolor{red}{As widely known, Besov spaces, $\left\lbrace \left(\mathcal{B}_{p,q}^{r}, \left\| \cdot \right\|_{p,q}^{r} \right), \ r \in \mathbb{R},~1 \leq p,q \leq \infty \right\rbrace$ can be interpreted as interpolated function spaces  between Sobolev spaces with fractional orders (which, in turn, constitute interpolated function spaces between Sobolev spaces with integer orders). We refer to the reader to Section 2.2 in the Supplementary Material, where the formal definition of Besov spaces is established, and their close relationship with Sobolev spaces is discussed.}
 
\textcolor{red}{The choice of Besov spaces is motivated by their ease of being represented in terms of}  the wavelet transform (see, e.g., \cite{Triebel83}). 
Specifically, for every $f\in \mathcal{B}_{p,q}^{r},$
\begin{equation}
\left\| f \right\|_{p,q}^{r} \equiv \left\|  \varphi_{J} \ast f \right\|_p + \left[ \displaystyle \sum_{j=J}^{\textcolor{red}{K}} \left(2^{j r}  \left\| \psi_j\ast f \right\|_p \right)^{q} \right]^{1/q} < \infty,
\label{normbesovspace} 
\end{equation}
\noindent where \textcolor{red}{$\varphi_J = \left\lbrace \varphi_{J,k}, \ k=0,1,\ldots,2^{J} - 1 \right\rbrace$ and $\psi_j = \left\lbrace \psi_{j,k}, \ k=0,1,\ldots,2^{j} \right\rbrace $, for each $j=J,J+1,\ldots,K$, denote the father and mother wavelets, and their translations and dilations, providing a Multiresolution Analysis  (MRA) of a suitable space of square-integrable functions. More details about MRA can be checked in Section 2.1 in the Supplementary Material; see also \cite{Daubechies92}}. Here, \textcolor{red}{$J$ is the primary resolution level, being required that $2^J \geq 2^{\left(\lceil r \rceil + 1 \right)}$ (see \cite{Angelinietal03}), and} $K$ defines the last  resolution level considered in the   
finite--dimensional wavelet \textcolor{red}{approximation.}

\medskip

\textcolor{red}{According to embeddings established in \cite[Section 6]{RuizAlvarez18_JMVA}, the simulation study undertaken in Section \ref{sec:4} below, the following function spaces  will be considered:}
\begin{eqnarray}
\overline{B} &=& \left[ \mathcal{B}_{\infty,\infty}^{0}  ([0,1]) \right]^{b+1};\quad \widetilde{\overline{H}} =  \left[ H_{2}^{-\beta} ([0,1]) \right]^{b+1}=\left[ \mathcal{B}_{2,2}^{-\beta }([0,1])\right]^{b+1}\nonumber\\
\overline{H} &=&  \left[L^2  ([0,1])\right]^{b+1};\quad 
\mathcal{H}(\overline{X})= \prod_{i=1}^{b+1}H^{\gamma_{i}}_{2}([0,1])=\prod_{i=1}^{b+1}\mathcal{B}_{2,2}^{\gamma_{i} }([0,1])\nonumber\\
\overline{B}^{\star } &=& \left[ \mathcal{B}_{1,1}^{0}  ([0,1]) \right]^{b+1};\quad \widetilde{\overline{H}}^{\star } =  \left[ H_{2}^{\beta} ([0,1]) \right]^{b+1}=\left[ \mathcal{B}_{2,2}^{\beta }([0,1])\right]^{b+1}
\nonumber\\
\overline{H}^{\star } &=&  \left[L^2  ([0,1])\right]^{b+1};\quad 
[\mathcal{H}(\overline{X})]^{\star }= \prod_{i=1}^{b+1}H^{-\gamma_{i}}_{2}([0,1])=\prod_{i=1}^{b+1}\mathcal{B}_{2,2}^{-\gamma_{i} }([0,1]),\nonumber\\
\label{eqfss}
\end{eqnarray}
\noindent where the parameters \textcolor{red}{$\{\gamma_{i}, \ i=1,\dots,b+1 \}$ (see Section \ref{sec:3b} below)}  reflect the second--order local regularity  of the  functional random components of $\overline{X}=\{\overline{X}_{n},\ n\in \mathbb{Z}\}$ in equation (\ref{eq_13}).  
 From embedding theorems between Besov spaces, the following continuous inclusions hold (see \cite{Triebel83}):
\begin{equation}\mathcal{H}(\overline{X})\hookrightarrow \widetilde{\overline{H}}^{\star} \hookrightarrow  \overline{B}^{\star}  \hookrightarrow   \overline{H} \hookrightarrow \overline{B}  \hookrightarrow 
 \widetilde{\overline{H}}\hookrightarrow [\mathcal{H}(\overline{X})]^{\star }, \label{eq_embed_simulation_2}
\end{equation}
\noindent
 for $\gamma_{i} >2\beta >1,$ $i=1,\dots,b+1.$  The $\overline{B}$  and $\overline{B}^{\star }$ norms are then computed from the following identities:  For every  $\overline{f} = \left(f; f_{1},\dots,f_{b} \right),\ \overline{g} = \left(g; g_{1},\dots, g_{b} \right)\in \overline{B}\subset \widetilde{\overline{H}},$
 
\begin{eqnarray}
\left\| \overline{f} \right\|_{\overline{B}} &=&  \displaystyle \sup_{j\geq J}\sup_{k =0,\dots, 2^{j}-1} \sup\left(\left| \alpha_{J,k}^{f}\right|,\left| \beta_{j,k}^{f}\right|,
\sup_{i =1,\dots, b}\left| \alpha_{J,k}^{f_{i}}\right|, \sup_{i =1,\dots, b}\left| \beta_{j,k}^{f_{i}}\right|\right)\nonumber\\
\nonumber \\
\left\| \overline{g} \right\|_{\textcolor{blue}{\overline{B}^{\star}}}^{2} &=& \left[\displaystyle \sum_{k=0}^{2^J-1}  \left| \alpha_{J,k}^{g}\right| + \displaystyle \sum_{j=J}^{K} \displaystyle \sum_{k=0}^{2^j - 1} \left| \beta_{j,k}^{g}\right| \right] +  \displaystyle   \left[\displaystyle \sum_{k=0}^{2^J-1} \sum_{i=1}^{b} \left| \alpha_{J,k}^{g_{i}}\right| + \displaystyle \sum_{j=J}^{K} \displaystyle \sum_{k=0}^{2^j - 1}\sum_{i=1}^{b} \left| \beta_{j,k}^{g_{i}}\right| \right], \nonumber \\ \label{fdecomp}
\end{eqnarray}
\noindent where, for $f\in B,$ and $g\in B^{\star},$
\begin{eqnarray}
\left\| f \right\|_{B} &=&  \displaystyle \sup \left\lbrace \left| \alpha_{J,k}^{f}\right|,~k =0,\ldots,2^{J}-1; ~\left| \beta_{j,k}^{f}\right|,~k=0,\ldots,2^{j}-1,~j=J,\ldots,K \right\rbrace, \nonumber  \\
\left\| g \right\|_{\textcolor{blue}{B^{\star}}} &=&  \displaystyle \sum_{k =0}^{2^{J}-1} \left|  \alpha_{J,k}^{g}\right| + \displaystyle \sum_{j=J}^{K} \displaystyle \sum_{k =0}^{2^{j}-1} \left| \beta_{j,k}^{g}\right|. \nonumber
\end{eqnarray}
 
\textcolor{red}{As usual, a proper representation in terms of wavelets  closely depends  on tuning parameters involved in equation  (\ref{fdecomp}). Since $r=0$, and $2^J \geq 2^{\left(\lceil r \rceil + 1 \right)}$ is required according to \cite{Angelinietal03},  $J=2$ is fixed, as the minimum number to be considered (the smaller value of $J$, the greater accuracy in the decomposition can be reached since more details can be captured). As deeply discussed in \cite[pp. 152--156]{AntoniadisSapatinas03}, the choice of the optimal $K$ can be just reduced to being such that $K < \log_2 (\sqrt{L})$, where $2^L$ denotes the number of grid points to be adopted for the generation of wavelets bases. In the numerical results displayed in Section \ref{sec:4},  Daubechies wavelets (with order $N = 10$; see \cite{Daubechies92} and Figure \ref{fig:2} below) 
with $L = 13$ have been considered (the maximum number of grid points to be considered under our computational restrictions), and $K = 6$ has been selected  as the optimal last resolution level by a classical cross-validation procedure. As shown in Sections \ref{sec:41}--\ref{sec:42}, note that trajectories can be generated with an alternative discretization step, just smoothing wavelet bases.}

\begin{figure}[H]
\begin{center}
 \includegraphics[width=0.9\textwidth]{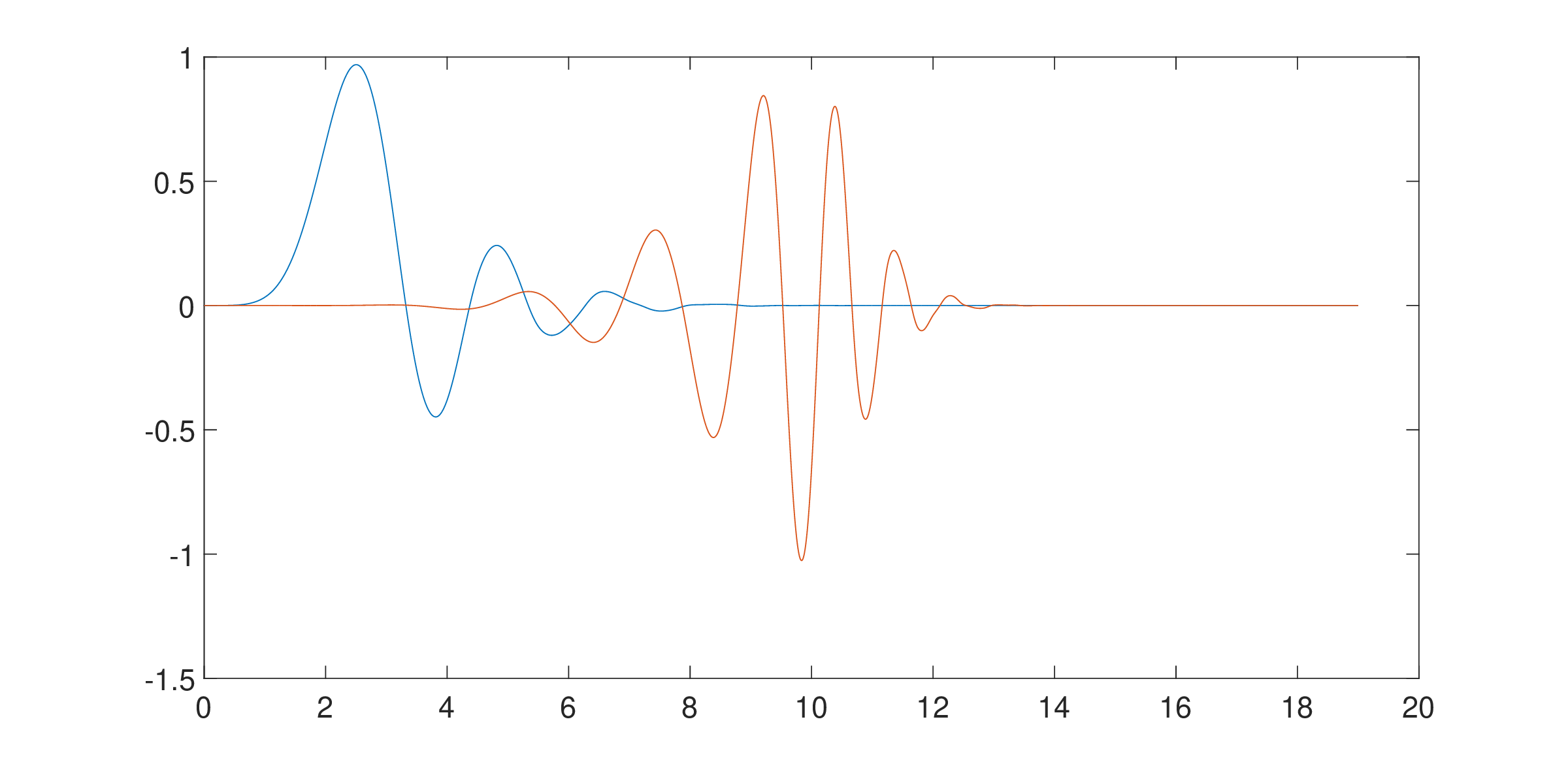}
  \vspace{-0.2cm}
 \caption{{\small  Father (blue) and mother (red) Daubechies wavelets with order $N=10$.}}
 \label{fig:2}
\end{center}
\end{figure}

\subsection{\textcolor{red}{General model and theoretical assumptions}}
 \label{sec:3b}
 
\textcolor{red}{Without loss of generality, we will considered three exogenous $\mathcal{B}_{\infty, \infty}^{0} \left([0,1]\right)$-valued variables (i.e., $b = 3$). According to the theoretical model presented in (\ref{eq_13}), and considering the function spaces defined in (\ref{eqfss}),} the following  ARBX(1) process has been generated:

\begin{eqnarray}
\overline{X}_n  &=& \overline{\rho} \left( \overline{X}_{n-1} \right) + \overline{\varepsilon}_{n},\quad \textcolor{blue}{\overline{X}_n,~\overline{\varepsilon}_{n} \in \overline{B}}, \quad n \in \mathbb{Z}, \nonumber \\
\overline{X}_n &=& \begin{pmatrix} X_n \\ Z_{n+1,1} \\ Z_{n+1,2} \\  Z_{n+1,3} \end{pmatrix}, \quad \overline{\varepsilon}_{n}= \begin{pmatrix} \varepsilon_n \\ \eta_{n,1} \\ \eta_{n,2} \\  \eta_{n,3} \end{pmatrix},  \quad \overline{\rho} = \begin{pmatrix} \rho & a_1 & a_2 &  a_3 \\ \mathbf{0}_B & u_{1} & \mathbf{0}_{B}  & \mathbf{0}_{B}  \\   \mathbf{0}_B & \mathbf{0}_{B}  & u_{2}  & \mathbf{0}_{B}  \\ \mathbf{0}_B & \mathbf{0}_{B}  & \mathbf{0}_{B}   & u_{3} \end{pmatrix}, \quad \textcolor{blue}{X_n \in \mathcal{B}_{\infty, \infty}^{0} \left([0,1]\right)}, \nonumber \\ \label{general_model}
\end{eqnarray}
\noindent \textcolor{red}{where $Z_{n,i},~\eta_{n,i}$ are also valued in $B = \mathcal{B}_{\infty, \infty}^{0} \left([0,1]\right)$, for each $n \in \mathbb{Z}$ and $i=1,2,3$.}

\medskip

\textcolor{red}{As discussed in \cite[Section 6]{RuizAlvarez18_JMVA}, and from the embeddings results in \cite{Triebel83},
$$\mathcal{H}(X) = H_{2}^{\gamma} \left([0,1]\right) \hookrightarrow H_{2}^{\beta} \left([0,1]\right) \hookrightarrow \mathcal{B}_{1,1}^{0} \left([0,1]\right) \hookrightarrow H = L^2 \left([0,1]\right) \hookrightarrow \mathcal{B}_{\infty,\infty}^{0} \left([0,1]\right),$$
\noindent satisfying \textbf{Assumptions A4--A5} for the case $b=1$, as long as $\gamma > 2 \beta > 1$, being $\mathcal{H}(X)$ the Reproducing Kernel Hilbert space generated by the covariance operator $C = \left(I - \Delta \right)^{-\gamma}$ (the $2\gamma/\beta$ power of the Bessel potential of order $\beta$ restricted to $L^2 \left([0,1]\right)$. Since the Cartesian product preserves the continuous embeddings, \textbf{Assumptions A4--A5} are directly verified as long as $$\beta > 1/2, \quad \gamma_{i} > 2\beta, \quad i=1,2,3,4.$$}

\textcolor{red}{Note that, since $\widetilde{\overline{H}}^{\ast} = \left[ H_{2}^{\beta} \left([0,1]\right) \right]^{4}$, the smaller is the parameter $\beta$, the smaller local regularity is observed in our curves. Henceforth, $\beta = 3/5 > 1/2$ is fixed to illustrate our framework in one of the
most irregular scenarios. Concerning $\left\lbrace \gamma_i, \ i=1,2,3,4 \right\rbrace$, we exposed that $\gamma_i > 2 \beta$ is just required, for each $i=1,2,3,4$. See Remark \ref{rem:gamma} below.}

\medskip

\begin{remark}
\label{rem:gamma}

\textcolor{red}{Since $ \left\lbrace X_n, \ n \in \mathbb{Z} \right\rbrace$ constitutes our endogenous variable of interest to be predicted, for simplicity, it seems reasonable to suppose that $\gamma_1 < \gamma_i$, for each $i=2,3,4$; that is, exogenous functional random variables (auxiliary external information) are rather regular than the exogenous variables to be predicted. Hence, the following parametric family of values, in terms of monotonic increasing functions, will be considered in the numerical results in Section \ref{sec:4}:
\begin{equation}
\gamma^{1} = \left\lbrace \gamma_i = 2 \beta + \frac{i}{10}, \ i=1,\ldots,b+1 \right\rbrace, \quad \gamma^{2} = \left\lbrace  \gamma_i = 2 \beta + \log_{10} \left(i+1 \right), \ i=1,\ldots,b+1 \right\rbrace. \label{param_gamma}
\end{equation}}

\textcolor{red}{Note that $\gamma^{2}$ will provide more regular curves. From a theoretical point of view, note that each parametric family $$\gamma = \left\lbrace \gamma_i = 2 \beta + f(i), \ i=1,\ldots,b \right\rbrace, \quad f(i) > 0,~i=1,\ldots,b+1,$$ could be adopted.}
  \end{remark}

\subsubsection{Covariance structures}

 \textcolor{red}{In relation to the covariance structure of the model in (\ref{general_model}), and in order to ensure the first part of \textbf{Assumption A1}},  a truncated version, on a multidimensional finite interval,  of the following multivariate infinite-dimensional Gaussian measures  in $\overline{H}=H^{4}$ \textcolor{red}{will be} generated:
 \begin{equation}
\overline{X}_0 = \begin{pmatrix} X_0 \\ Z_{1,1} \\ Z_{1,2} \\  Z_{1,3} \end{pmatrix}\sim \mathcal{N}\left(\mathbf{0},\overline{C}\right), \quad \overline{\varepsilon}_{0}= \begin{pmatrix} \varepsilon_0 \\ \eta_{0,1} \\ \eta_{0,2} \\  \eta_{0,3} \end{pmatrix}\sim \mathcal{N}\left(\mathbf{0},\overline{C}_{\boldsymbol{\eta}}\right),
\label{indh}\end{equation}
 
\noindent where
 \begin{eqnarray}
\overline{C} &=& \begin{pmatrix} C_{X_{0},X_{0}} & C_{X_{0},Z_{1,1}} & C_{X_{0},Z_{1,2}} & C_{X_{0},Z_{1,3}} \\ C_{Z_{1,1},X_{0}} & C_{Z_{1,1},Z_{1,1}} & C_{Z_{1,1},Z_{1,2}} & C_{Z_{1,1},Z_{1,3}}\\ C_{Z_{1,2},X_{0}} & C_{Z_{1,2},Z_{1,1}} & C_{Z_{1,2},Z_{1,2}} & C_{Z_{1,2},Z_{1,3}}  \\ C_{Z_{1,3},X_{0}} & C_{Z_{1,3},Z_{1,1}} & C_{Z_{1,3},Z_{1,2}} & C_{Z_{1,3},Z_{1,3}}  \end{pmatrix},\label{c14} 
\end{eqnarray}
\noindent and 
\begin{eqnarray}
\overline{C}_{\boldsymbol{\eta}} &=& \begin{pmatrix} C_{\varepsilon_{0},\varepsilon_{0}} & C_{\varepsilon_{0},\eta_{0,1}} & C_{\varepsilon_{0},\eta_{0,2}} & C_{\varepsilon_{0},\eta_{0,3}} \\ C_{\eta_{0,1},\varepsilon_{0}} & C_{\eta_{0,1},\eta_{0,1}} & C_{\eta_{0,1},\eta_{0,2}} & C_{\eta_{0,1},\eta_{0,3}}\\ C_{\eta_{0,2},\varepsilon_{0}} & C_{\eta_{0,2},\eta_{0,1}} & C_{\eta_{0,2},\eta_{0,2}} & C_{\eta_{0,2},\eta_{0,3}}  \\ C_{\eta_{0,3},\varepsilon_{0}} & C_{\eta_{0,3},\eta_{0,1}} & C_{\eta_{0,3},\eta_{0,2}} & C_{\eta_{0,3},\eta_{0,3}}  \end{pmatrix}.\label{c14bb} 
\end{eqnarray}

In equation (\ref{c14}), we have denoted
\begin{equation}
C_{X_{0},X_{0}} =  \left(I - \Delta \right)^{-\gamma_1}, \quad C_{Z_{1,i},Z_{1,i}} = {\rm E} [Z_{1,i} \otimes Z_{1,i}] =  \left(I - \Delta \right)^{-\gamma_i},~i=1,2,3,\label{eqfco}
\end{equation}
 \noindent with $\left(I - \Delta \right)^{-\gamma}$ being the Bessel potential of order $2\gamma,$ and  $C_{Z_{1,i},X_{0}} = {\rm E} [Z_{1,i} \otimes X_0]$ and $C_{X_{0},Z_{1,i}} = {\rm E} [X_0 \otimes Z_{1,i}]$,  for  $i,l=1,2,3.$  
 The functional entries of (\ref{c14}) are now  explicitly defined, ensuring, in particular, the  one--dimensionality of their eigenspaces, in order to get  the second part of  \textbf{Assumption A1} to hold.      \textcolor{blue}{Furthermore}, for  every $f \in L^{2}([0,1]),$ and for $i,l=1,2,3,$
\begin{eqnarray}
C_{X_{0},X_{0}} \left( f \right)  &=& \displaystyle \sum_{j=1}^{\infty} (1 + j)^{-\gamma_1} \langle \phi_j, f \rangle_{H}  \phi_j, \nonumber\\  C_{X_{0},Z_{1,i}} \left( f \right)   &=&  \displaystyle \sum_{j=1}^{\infty} (1+j)^{-\frac{\gamma_1+\gamma_{i+1}}{2}}  \langle \phi_j, f \rangle_{H}  \phi_j, \nonumber \\
  C_{Z_{1,i},X_{0}} \left( f \right) &=& \displaystyle \sum_{j=1}^{\infty} (1+j)^{-\frac{\gamma_1+\gamma_{i+1}}{2}}   \langle \phi_j, f \rangle_{H}  \phi_j, \nonumber \\   C_{Z_{1,l},Z_{1,i}} \left( f \right)  &=& \displaystyle \sum_{j=1}^{\infty} (1+j)^{-\frac{\gamma_{l}+\gamma_{i+1}}{2}}  \langle \phi_j, f \rangle_{H} \phi_j , \nonumber\\  \label{covar_oper} 
\end{eqnarray}

\textcolor{red}{Henceforth, the following basis will be adopted, in the decompositions exposed in (\ref{covar_oper}):}
 \begin{equation}
\phi_j \left( x \right) = \sqrt{\frac{2}{b-a}} {\rm sin}\left(\frac{\pi j x}{b-a}\right), \quad j \geq 1, \quad  x \in \left[a,b\right],\quad a=0,\ b=1. \label{sl}
\end{equation}

 \textcolor{red}{That is, the smoother functional values of the  exogenous  random variables are extended to the space $H=L^{2}([0,1]),$  by projection   into the elements of the  basis in (\ref{sl}). Adopting as an example the parametric family $\gamma^{1}$ displayed in (\ref{param_gamma}), the functional entries  $C_{X_{0},X_{0}}$ and  $C_{Z_{0,3},Z_{0,3}}$ are represented in Figure,   \ref{fig:3b} below.}

\begin{figure}[H]
\hspace{-0.8cm} \includegraphics[width=0.58\textwidth]{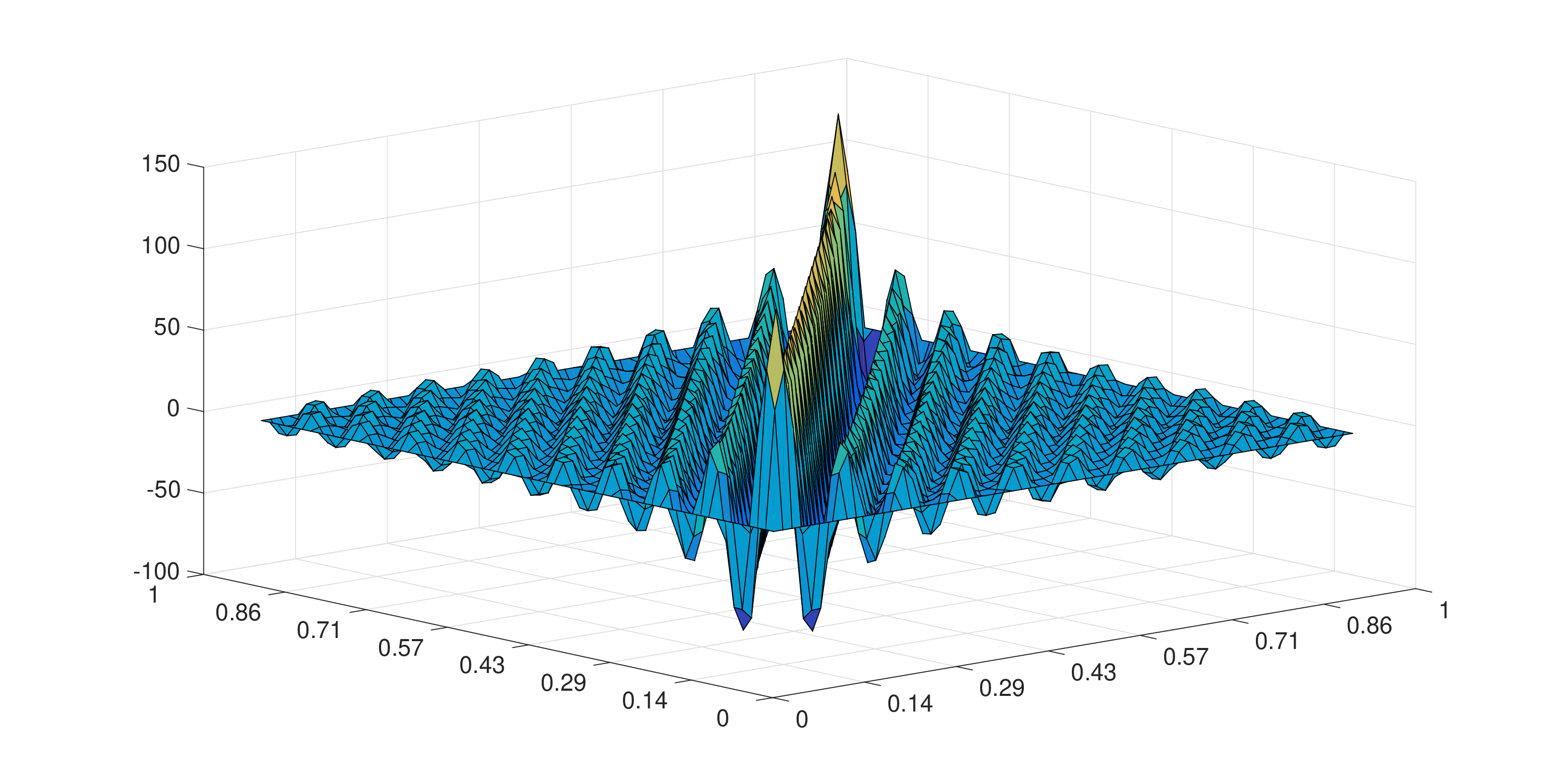}
\hspace{-0.65cm} \includegraphics[width=0.58\textwidth]{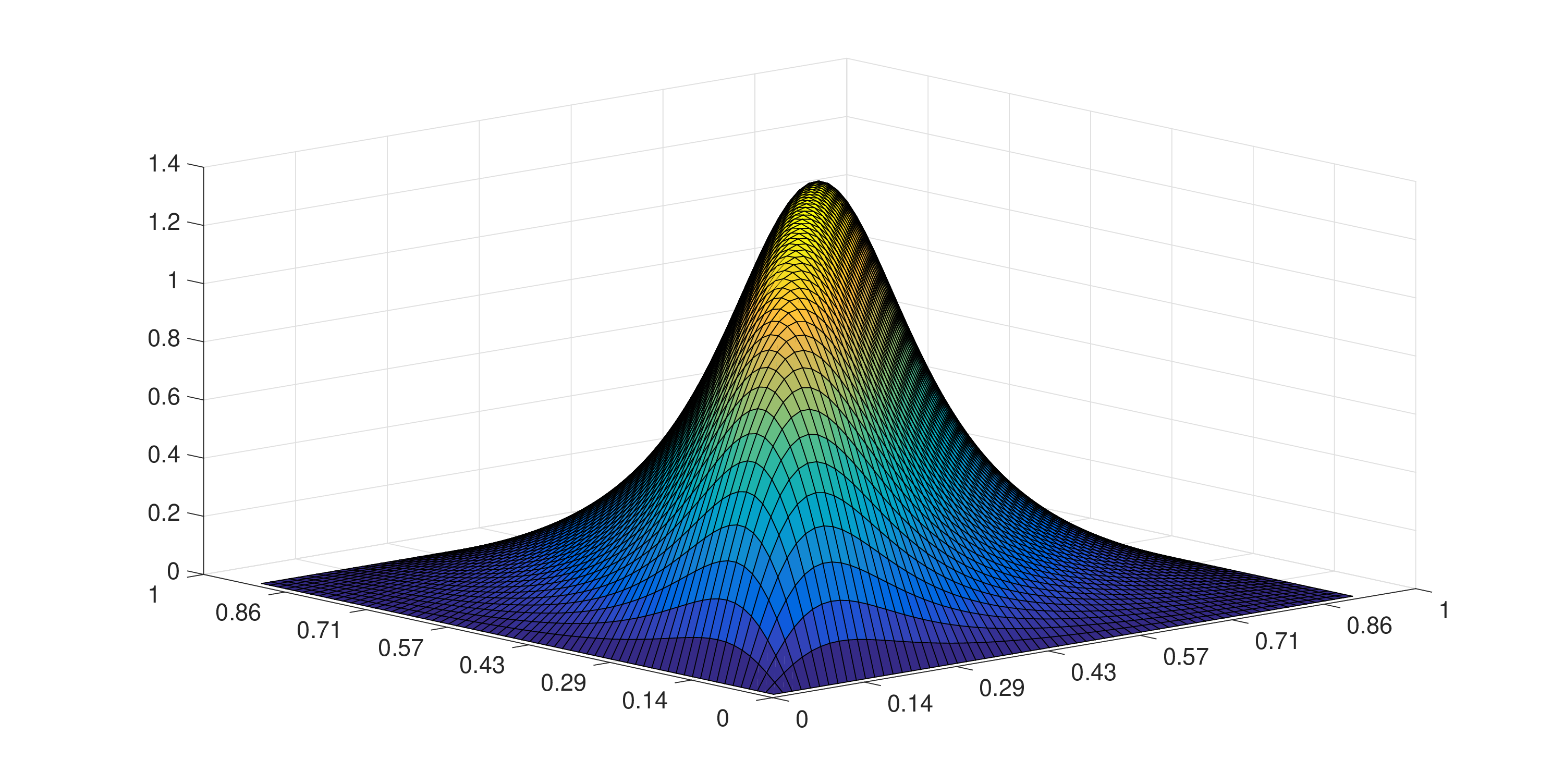}
  \vspace{-0.2cm}
 \caption{{\small  Covariance kernels defining $C_{X_{0},X_{0}}$ (on left) and $C_{Z_{1,3},Z_{1,3}}$ (on right), generated in terms of $\gamma_1 = 2\beta + 1/10$ and $\gamma_4 = 2\beta + 4/10$ with $\beta = 3/5$, respectively, plotted with a discretization step size $\Delta h = 0.0159$.}}
 \label{fig:3b}
\end{figure}

\medskip

\textcolor{red}{In the same way,  the functional entries of 
 (\ref{c14bb}) are explicitly defined as follows:}
 \begin{eqnarray} && C_{\varepsilon_{0},\varepsilon_{0}} = {\rm E} \left[ \varepsilon_0 \otimes \varepsilon_0 \right],\quad  C_{\eta_{0,i},\eta_{0,l}} =  {\rm E} \left[ \eta_{0,i} \otimes \eta_{0,l} \right], \ i,l=1,2,3 \nonumber\\  && C_{\eta_{0,i},\varepsilon_{0}}={\rm E} \left[ \eta_{0,i} \otimes \varepsilon_0 \right],\quad C_{\varepsilon_{0},\eta_{0,i}}={\rm E} \left[   \varepsilon_0 \otimes\eta_{0,i}\right],\ i=1,2,3.\label{feicm}\end{eqnarray}

\textcolor{red}{In particular, and for simplifications purposes}, we have considered, in the generations,  $C_{\eta_{0,i},\eta_{0,l}}=C_{\varepsilon_{0},\eta_{0,i}}= \mathbf{0}_{H},$ $i,l=1,2,3,$ $i\neq l.$ The following identities characterise   the diagonal functional entries of $\overline{C}_{\boldsymbol{\eta}},$ in terms of the elements of the basis $\{ \phi_{j},\ j\geq 1\}$ introduced in equation (\ref{sl}): For $i=1,2,3,$ and  $j,h\geq 1,$ 
\begin{eqnarray}
\langle C_{\varepsilon_{0},\varepsilon_{0}}(\phi_j), \phi_h \rangle_{H} &=& \begin{cases}
C_{X_{0}X_{0}}(\phi_{j})(\phi_{j}) (1-[\rho (\phi_{j})(\phi_{j})]^{2}) \quad & j =h, \\
e^{- \left| j - h \right|^2 / W^2} \quad & j \neq h  
\end{cases}, \nonumber \\
\langle C_{\eta_{0,i},\eta_{0,i}}(\phi_j), \phi_h \rangle_{H} &=& \begin{cases}
C_{Z_{0,i},Z_{0,i}}(\phi_{j})(\phi_{j}) \left(1-[u_{i}(\phi_{j})(\phi_{j})]^{2}\right) \quad & j =h \\
e^{- \left| j - h \right|^3 / W^2} \quad & j \neq h  
\end{cases}, \nonumber
\end{eqnarray}
\noindent where, 
for $i=1,2,3,$ and $j,h\geq 1,$ considering  $W = 0.4,$
\begin{eqnarray}
\rho(\phi_j)(\phi_h)=\langle \rho (\phi_j), \phi_h \rangle_{H} &=& \begin{cases}
(1+j)^{-1.5} \quad & j =h, \\
e^{- \left| j - h \right| / W} \quad & j \neq h  
\end{cases}, \nonumber \\
u_{i}(\phi_j)(\phi_h)=\langle u_i (\phi_j), \phi_h \rangle_{H} &=& \begin{cases}
(1+j)^{-(3 + 0.5 i)} \quad & j =h, \\
e^{- \left| j - h \right|^2 / W} \quad & j \neq h.  
\end{cases}\end{eqnarray}

Furthermore, for $i=1,2,3,$ and   
$j,h\geq 1,$   
\begin{equation}
\langle a_i(\phi_j), \phi_h \rangle_{H} = \begin{cases}
(1+j)^{-(4 + 0.5 i)} \quad  j =h, \\
e^{- \left| j - h \right|^3 / W} \quad  j \neq h  
\end{cases}. \label{acoex}
\end{equation}

\section{\textcolor{red}{Simulation study}} 
\label{sec:4}

  \textcolor{red}{There is a twofold objective for the numerical results displayed in this Section: firstly, to illustrate the strong consistency of the ARBX(1) plug-in predictor for an increasing sequence of sample sizes and, secondly, to explore whether our ARBX(1) prediction methodology is sensitive to the number of grid points for a decreasing (to zero) sequence of discretization steps.}
  
  \medskip
  
  As given in Section \ref{sec:2}, the  componentwise estimator   (\ref{estimator}) of $\overline{\rho}$  is \textcolor{blue}{strongly} consistent in $\mathcal{L}(\overline{B}),$ under the formulated \textbf{Assumptions A1--A5}, and the conditions in \textcolor{blue}{\cite[Lemma 8--9]{RuizAlvarez18_JMVA}}. \textcolor{red}{The general model proposed in  Section
 \ref{sec:3} has been demonstrated to satisfy
  the theoretical conditions included in \textbf{Assumption A1} and \textbf{Assumptions A4--A5}. Regarding empirical conditions displayed in \textbf{Assumptions A2--A3}, for a given functional sample size $n$, in the following, we will consider  $k_{n}=[\ln(n)]^{-},$ where $[\cdot]^{-}$ denotes the integer part function. This truncation parameter ensures that, for all the sample sizes studied, the empirical eigenvalue $C_{n,k_{n}}$ is positive, and therefore, \textbf{Assumption A2} is satisfied. See Figures \ref{fig:2h}--\ref{fig:3} in the Appendix for checking how \textbf{Assumption A3} and extra condition required in \cite[Theorem 1]{RuizAlvarez18_JMVA} are respectively satisfied.}

\subsection{\textcolor{blue}{Strong} consistency of the ARBX(1) plug--in predictor}
\label{sec:41}

\textcolor{red}{As commented, and aimed at illustrating  the behaviour of our plug--in predictor for very large sample sizes, let us consider in this Section the following increasing sequence of sample sizes:
$$n =  \left[n_1, \ldots, n_9 \right] = \left[1500, 2500, 5000, 15000, 25000, 50000, 75000, 100000, 130000\right]$$}

\textcolor{red}{Some trajectories of $X_n^{\star }\in \widetilde{H}^{\star}=\mathcal{B}^{\beta}_{2,2}([0,1])$ have been plotted in Figure \ref{fig:5} below, adopting a discretization step $\Delta h = 0.0159$, just  smoothing with  \textit{cubicspline} option   of \texttt{fit.m} \texttt{MatLab} function. } 
 
  \begin{figure}[H]
 \centering
\includegraphics[width=0.92\textwidth]{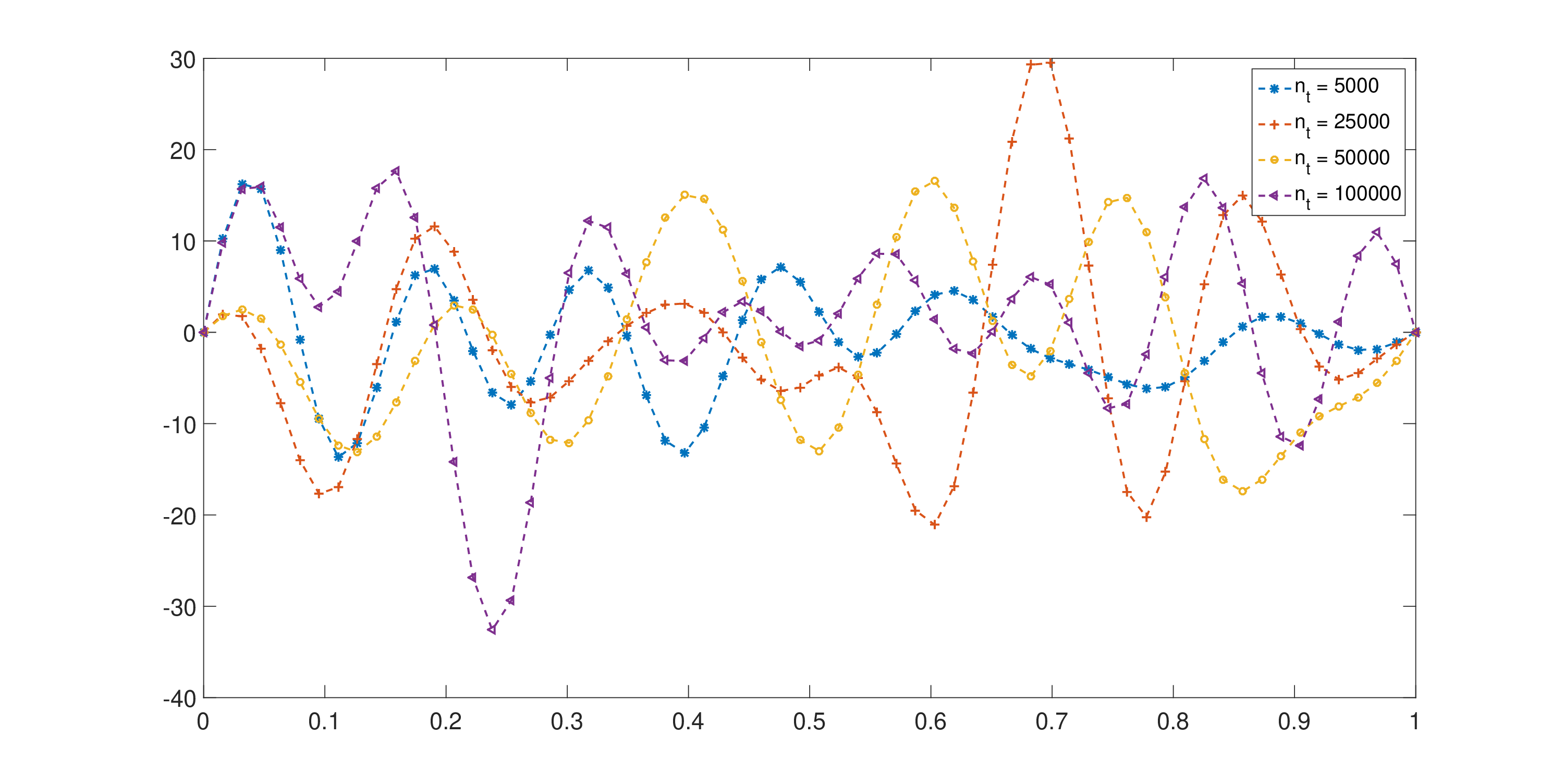}
  \vspace{-0.2cm}
 \caption{{\small    Functional values of $X_n^{\star}$ at  $n=5000,25000,50000,100000$, with discretization step size $\Delta h = 0.0159$.}}
 \label{fig:5}
\end{figure}

\textcolor{red}{As noted, we have adopted $k_{n}=[\ln(n)]^{-}$ as truncation parameter, such that,} according to \cite[Theorem 1]{RuizAlvarez18_JMVA}, 
the strong-consistency of the componentwise estimator (\ref{estimator})   in $\mathcal{L}(\overline{B})$ follows when   \begin{equation}k_{n}C_{k_{n}}^{-1}\sum_{j=1}^{k_{n}} a_j =o\left(\sqrt{n/\ln(n)}\right),\quad n\to \infty,\label{sfcth1}
\end{equation}
\noindent where $$a_1=\frac{2\sqrt{2}}{C_{1}-C_{2}};\ a_{j}=2\sqrt{2}\max\left(\frac{1}{C_{j-1}-C_{j}},\frac{1}{C_{j}-C_{j+1}}\right),\quad j\geq 2.$$
 \noindent As before, \textcolor{blue}{$\{C_{j}, \ j\geq 1 \}$} denotes the system of eigenvalues of the extended matrix autocovariance operator $\overline{C}$ in (\ref{c14}). We can observe in Figure \ref{fig:3} \textcolor{red}{in the Appendix} that condition (\ref{sfcth1}) is satisfied.

 \medskip
 
\textcolor{blue}{For numerically proving the strong consistency of the ARBX(1) plug--in predictor, note that} the upper bound derived in \textcolor{blue}{\cite[Theorem 1 and Corollary 1]{RuizAlvarez18_JMVA}}, for the $\overline{B}$--norm of the  functional error, $\left\| \overline{\rho} (\overline{X}_{n}) - \widehat{\overline{X}}_{n} \right\|_{\overline{B}},$ ensuring strong consistency of the plug--in predictor in $\overline{B},$ is given by 
 \begin{equation}
\mathcal{M}(\omega )\xi_{n} = \mathcal{M}(\omega )\exp \left(\frac{-n}{C_{k_{n}}^{-2} k_{n}^{2} \left(\displaystyle \sum_{j=1}^{k_{n}} a_j \right)^2} \right), \label{upper_bound}
\end{equation}
\noindent where  $\mathcal{M}(\omega )=\|\overline{X}_{0}(\omega )\|_{\overline{B}},$ $\omega \in \Omega,$ and  for each functional sample size $n,$ as before,
  $C_{k_{n}}$ is  the $k_{n}$--th eigenvalue of the extended autocovariance operator $\overline{C}$  in (\ref{c14}). \textcolor{blue}{Hence,}  Table \ref{tab:1}  reflects the \textcolor{red}{percentage of simulations in which} the  error $\overline{B}$--norm \textcolor{blue}{is} greater than the  upper bound  \textcolor{blue}{in} (\ref{upper_bound})\textcolor{red}{, for each one of the parametric families of $\gamma$ defined in (\ref{param_gamma}).}

  \renewcommand{\arraystretch}{1.4}
  
 \begin{table}[H]
\caption{{\small Percentage  of simulations from the  $200$ generations of each sample size, where  the error $\overline{B}$--norm is larger than   the upper bound  (\ref{upper_bound}).  The sample sizes $n = 1500, 2500, 5000, 15000, 25000, 50000, 75000, 100000, 130000$ have been tested.  The truncation rule $k_{n} = \left[ \ln(n) \right]^{-}$  has been adopted. \textcolor{red}{Two parametric families $\gamma^{1}$ and $\gamma^{2}$ have been adopted for the generation of covariance operators (see equation (\ref{param_gamma})).} }}
\centering
\begin{small}
\begin{tabular}{|c||c|c|}
 \hline
    $n_t$ & $\gamma^{1}$ & $\gamma^{2}$ \\
    \hline
    $n_1 = 1500$ & $11.5~\%$ ($\frac{23}{200}$) & $12~\%$ ($\frac{24}{200}$)  \\
    \hline
    $n_2 = 2500$ & $9.5~\%$ ($\frac{19}{200}$) & $9~\%$ ($\frac{18}{200}$)   \\
    \hline
	$n_3 = 5000$ & $8~\%$ ($\frac{16}{200}$) & $8.5~\%$ ($\frac{17}{200}$)  \\
    \hline
	$n_4 = 15000$ & $4.5~\%$ ($\frac{9}{200}$) & $4.5~\%$ ($\frac{9}{200}$)  \\
    \hline
	$n_5 = 25000$ & $3.5~\%$ ($\frac{7}{200}$) & $2.5~\%$ ($\frac{5}{200}$)  \\
    \hline
	$n_6 = 50000$ & $2.5~\%$ ($\frac{5}{200}$) & $1.5~\%$ ($\frac{3}{200}$)   \\
    \hline
	$n_7 = 75000$ & $2~\%$ ($\frac{4}{200}$) & $1~\%$ ($\frac{2}{200}$)   \\
    \hline
	$n_8 = 100000$ & $1~\%$ ($\frac{2}{200}$) & $0.5~\%$ ($\frac{1}{200}$)   \\
    \hline
	$n_9 = 130000$ & $0~\%$ ($\frac{0}{200}$) & $0~\%$ ($\frac{0}{200}$)  \\
    \hline    
\end{tabular} 
\end{small}
  \label{tab:1}
\end{table}

\subsection{\textcolor{red}{Asymptotic behaviour of discretely observed ARBX(1) processes}}
\label{sec:42}

\textcolor{red}{The previous subsection was intended to numerically illustrate the strong consistency of ARBX(1) plug--in predictor, focusing on its behaviour when $n \to \infty$, for a fixed discretization step in the generation of both trajectories and wavelets. In contrast, the main aim of this subsection is  to explore the sensitiveness of the above  ARBX(1) prediction methodology to  the discretization step size. Here, we provide the reader a brief numerical study about what is going on when sample sizes are not too large ($n$ does not tend to infinite) but the discretization step adopting for the generation of the trajectories tends to zero ($\Delta h \to 0$).}

\medskip

\textcolor{red}{For this purpose, the set $\left\lbrace \Delta h_j = \frac{1}{3^{2+ j}}, \ j=1,\ldots,7 \right\rbrace$ of decreasing (to zero) discretization steps  are analysed (see Figure \ref{fig:h} below): 
\begin{eqnarray}
 \Delta h_1 &=& 3.70 (10^{-2}), \quad \Delta h_2 = 1.23 (10^{-2}), \quad \Delta h_3 = 4.12 (10^{-3}), \quad \Delta h_4 = 1.37 (10^{-3}),\nonumber \\
\Delta h_5 &=& 4.57 (10^{-4}), \quad \Delta h_6 = 1.52 (10^{-4}), \quad \Delta h_7 = 5.08(10^{-5}), \nonumber 
\end{eqnarray}
\noindent such that $\left\lbrace 3^{2+ j} + 1, \ j=1,\ldots, 7 \right\rbrace$ grid points are respectively considered.}
 
 \vspace{-0.3cm}
   \begin{figure}[H]
 \centering
\includegraphics[width=0.95\textwidth]{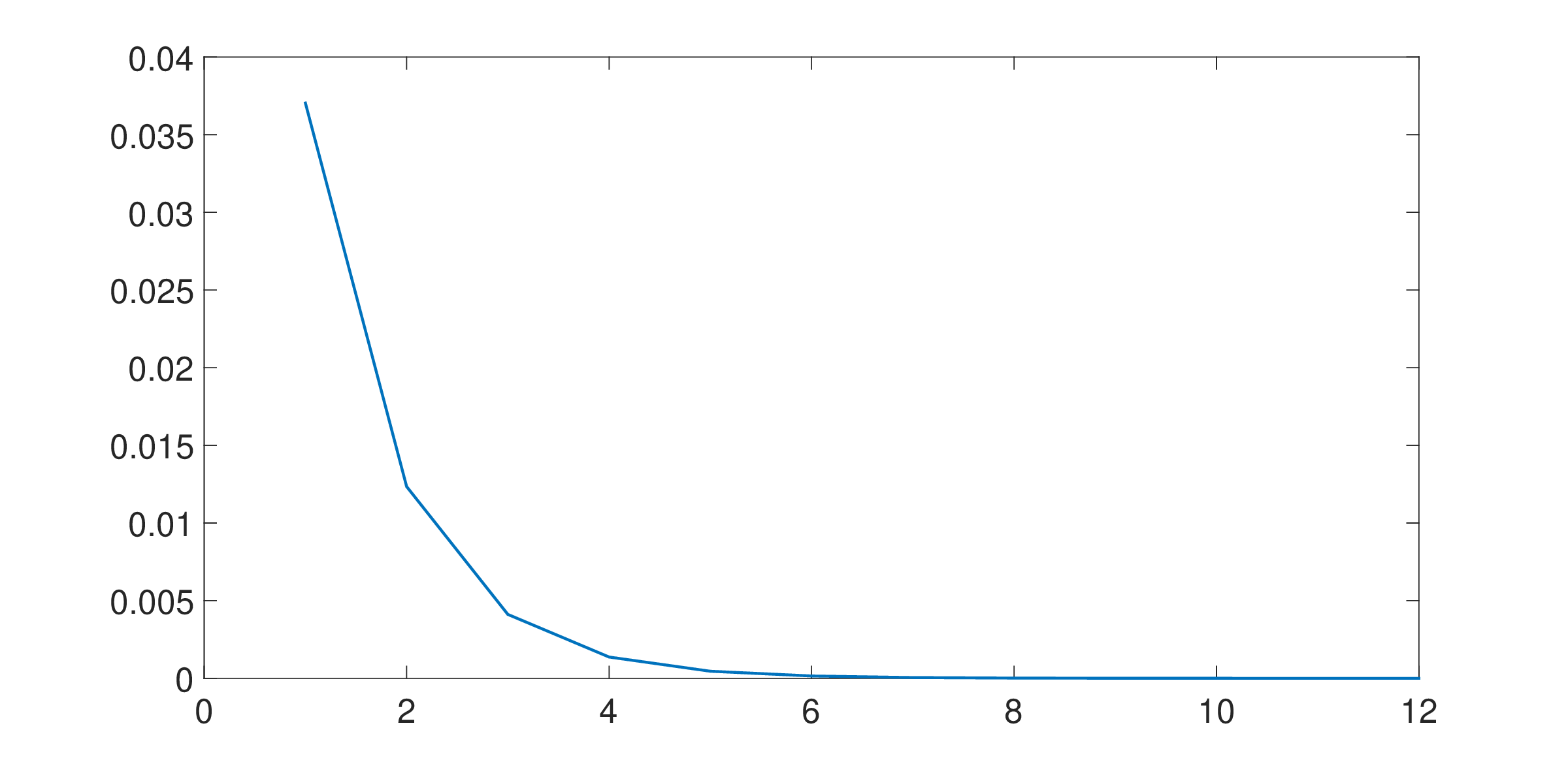}
  \vspace{-0.2cm}
 \caption{{\small  Discretization steps $\Delta h_j = \frac{1}{3^{2+ j}}$, for $j=1,\ldots,12$, displaying that $\Delta h_j \to 0$ as $j \to \infty$.}}
 \label{fig:h}
\end{figure}

\vspace{-0.35cm}
\textcolor{red}{Now, only  sample sizes $n= \left[n_1, n_2, n_3 \right] = [5000,15000,30000]$ have been considered. As below,  Table \ref{tab:2} reflects the percentages of simulations in which   the error $B$--norm  is larger than the  upper bound in (\ref{upper_bound}), for each discretization step size,  functional sample size  and parametric families $\gamma^{1}$ and $\gamma^{2}$.}

\setlength{\tabcolsep}{2.8pt}
  \renewcommand{\arraystretch}{1.6}

 \begin{table}[H]
\caption{{\small   \textcolor{red}{Percentage  of simulations from the  $200$ generations of each sample size, where  the error $\overline{B}$--norm is larger than   the upper bound  (\ref{upper_bound}).  The sample sizes $n = 5000, 15000, 30000$ have been tested.  The truncation rule $k_{n} = \left[ \ln(n) \right]^{-}$  has been adopted. Two parametric families $\gamma^{1}$ and $\gamma^{2}$ have been adopted for the generation of covariance operators (see equation (\ref{param_gamma})). Discretization steps $\left\lbrace \Delta h_j = 3^{-(2+j)}, \ j=1,\ldots, 7 \right\rbrace$ are adopted.}}}
\centering
\begin{small}
\begin{tabular}{|c||c|c|c||c|c|c|}
 \hline
 & \multicolumn{3}{c||}{$\gamma^{1}$} & \multicolumn{3}{c|}{$\gamma^{2}$} \\
 \hline \hline
  $n \longrightarrow$  & $5000$ & $15000$ & $30000$  & $ 5000$ & $ 15000$ & $30000$ \\
    \hline
    $\Delta h_1 $ & $12~\%$ ($\frac{24}{200}$) & $10~\%$ ($\frac{20}{200}$) & $7.5~\%$ ($\frac{15}{200}$) & $13.5~\%$ ($\frac{27}{200}$) & $10.5~\%$ ($\frac{21}{200}$) & $6.5~\%$ ($\frac{13}{200}$)  \\
    \hline
    $\Delta h_2  $ & $9~\%$ ($\frac{18}{200}$) & $5.5~\%$ ($\frac{11}{200}$) & $4~\%$ ($\frac{8}{200}$) & $9.5~\%$ ($\frac{19}{200}$) & $6~\%$ ($\frac{12}{200}$) & $4~\%$ ($\frac{8}{200}$)  \\
    \hline
    $\Delta h_3 $ & $7~\%$ ($\frac{14}{200}$) & $4~\%$ ($\frac{8}{200}$) & $3.5~\%$ ($\frac{7}{200}$) & $7.5~\%$ ($\frac{15}{200}$) & $4~\%$ ($\frac{8}{200}$) & $3~\%$ ($\frac{6}{200}$)  \\
    \hline
    $\Delta h_4 $ &  $5.5~\%$ ($\frac{11}{200}$) & $3.5~\%$ ($\frac{7}{200}$) & $2~\%$ ($\frac{4}{200}$) & $4.5~\%$ ($\frac{9}{200}$) & $2.5~\%$ ($\frac{5}{200}$) & $2~\%$ ($\frac{4}{200}$)   \\
    \hline
    $\Delta h_5  $ & $2.5~\%$ ($\frac{5}{200}$) & $1.5~\%$ ($\frac{3}{200}$) & $1~\%$ ($\frac{2}{200}$) & $1.5~\%$ ($\frac{3}{200}$) & $1~\%$ ($\frac{2}{200}$) & $1~\%$ ($\frac{2}{200}$)  \\
    \hline
    $\Delta h_6  $ & $1.5~\%$ ($\frac{3}{200}$) & $0.5~\%$ ($\frac{1}{200}$) & $0.5~\%$ ($\frac{1}{200}$) & $1~\%$ ($\frac{2}{200}$) & $0.5~\%$ ($\frac{1}{200}$) & $0~\%$ ($\frac{0}{200}$)  \\
    \hline 
    $\Delta h_7 $ & $1~\%$ ($\frac{2}{200}$) & $0.5~\%$ ($\frac{1}{200}$) & $0~\%$ ($\frac{0}{200}$) & $0.5~\%$ ($\frac{1}{200}$) & $0~\%$ ($\frac{0}{200}$) & $0~\%$ ($\frac{0}{200}$)  \\
    \hline    
\end{tabular} 
\end{small}
  \label{tab:2}
\end{table}

\textcolor{red}{In the light of the results displayed in Tables \ref{tab:1}--\ref{tab:2}, our ARBX(1) plug--in predictor is strongly consistent, either considering sample sizes tending to infinite or considering discretely observed ARBX(1) process adopting a discretization step converging to zero.}

\section{Real-data application: short--term forecasting of air pollutants}
\label{sec:5}

In this section, the performance of the ARBX(1) based prediction approach \textcolor{blue}{here} presented is illustrated in  a real--data example. Specifically, the short--term forecasting of daily average concentrations of atmospheric aerosol particles with diameters less than 10 $\mu m$, also known as PM$_{10}$ (coarse particles), is achieved from a functional perspective. The importance of the accurate  forecasting  of this kind of particles relies on the fact that they  are inhalable atmospheric pollution particles, which impact the public health. \textcolor{red}{Following the suggestions by the World Health Organization, the European Union developed in 2008 (in particular, directive 2008/50/EU) a complete legislative package, establishing health based standards for the levels of PM$_{10}$: daily mean concentration of PM$_{10}$ should not be greater than $50~\mu g~m^{-3}$ more than 35 days per year, neither the annual average of concentration of PM$_{10}$ shall not be greater than $40~\mu g~m^{-3}$.} However, this limit has been exceed during the last years in heavily industrialized areas, deriving in severe people's health problems. Therefore, PM$_{10}$ forecasting is crucial to adopting efficient public transport policies. The dataset is analysed  in Section \ref{sec:51}, while Section \ref{sec:52} describes  the  \textcolor{blue}{preprocessing} procedure required, before implementing our functional prediction methodology in Section \ref{sec:44}.

\subsection{\textcolor{blue}{Dataset description}}
\label{sec:51}

\begin{figure}[H]
  \centering
  \includegraphics[width=1.05\textwidth]{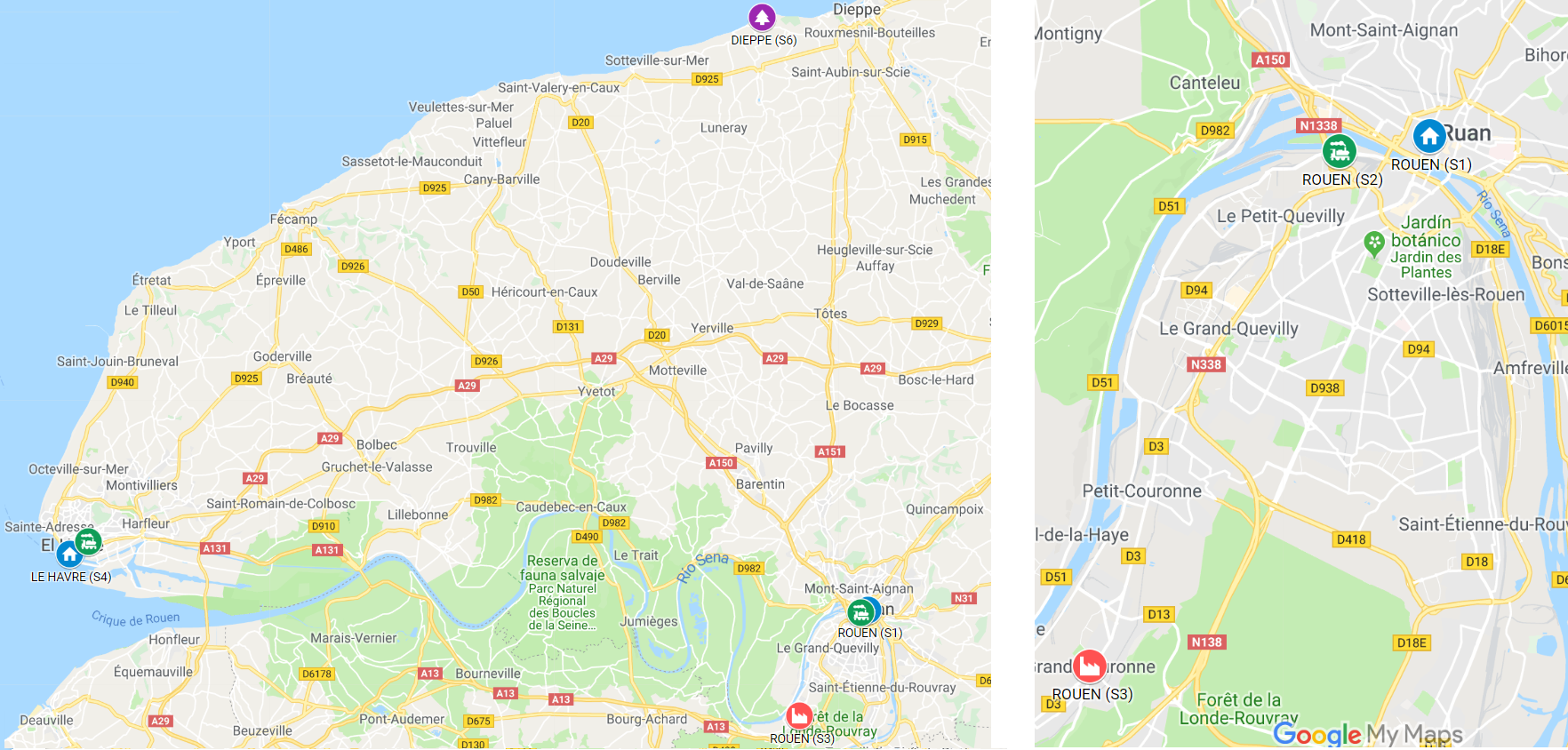}     
  \vspace{-0.18cm}
 \caption{{\small \textcolor{red}{On left: map (extracted from Google Maps) displaying the location of the six pollution monitoring stations analysed. On right: enlarged map  displaying the  stations near Rouen.} }}
 \label{fig:7}
\end{figure}

The dataset considered is comprised of daily average concentrations of PM$_{10}$, coming from hourly measurements, from  January 1, 2007 to  March 31,  2011,  collected by the air quality Normand (French) authority, known as Air Normand. This dataset is freely available in the website \url{http://www.atmonormandie.fr}. Specifically, we pay attention to $6$ of the $13$ fixed pollution monitoring stations network located throughout Haute--Normandie region, considered one of the most heavily industrialized areas in France (see locations in \textcolor{red}{the map displayed in Figure \ref{fig:7}}). 

In the following, \textcolor{red}{as shown in the map below,} these  monitoring stations will be denoted as $\left\lbrace S_c, \ c=1,\ldots,6 \right\rbrace$, \textcolor{blue}{selected with the aim of reflecting} a wide variety of scenarios, such that roadisdes \textcolor{red}{(stations S2 and S5)}, urban areas \textcolor{red}{(stations S1 and S4)}, industrial zones \textcolor{red}{(station S5)} and rural regions \textcolor{red}{(station S6)}. \textcolor{red}{As reflected in  Figure \ref{fig:9} below (see also descriptive statistics reflected in Table \ref{tab:3} in the Appendix)}, the monitoring station $S_6$ (\textcolor{red}{rural area}) has \textcolor{blue}{registered} the smallest PM$_{10}$ concentrations, while stations $S_2$ and $S_5$ (\textcolor{red}{roadsides}) \textcolor{blue}{display} the highest pollution levels. These stations also display the highest  variability, which seems logical since pollution levels in roadside are strongly dependent on \textcolor{blue}{traffic jams}.

 \vspace{-0.3cm}
\begin{figure}[H]
 \centering
  \includegraphics[width=0.9\textwidth]{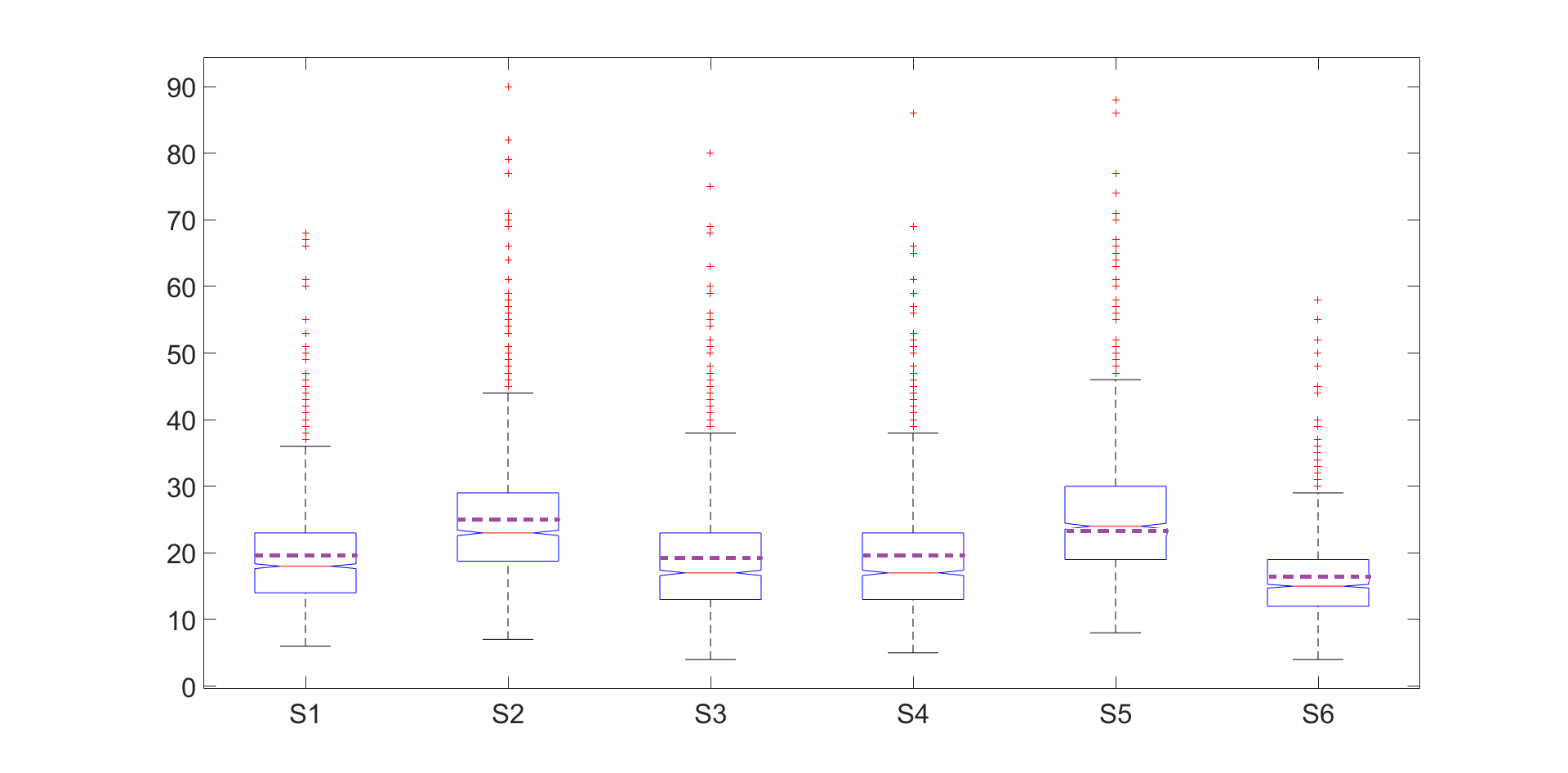}
 \vspace{-0.25cm}
 \caption{{\small \textcolor{red}{Boxplots on the samples of the  PM$_{10}$ concentrations ($\mu g~m^{-3}$). Purple dotted lines reflect the average concentrations, while red solid lines splitting the box reflect the medians. }}}
 \label{fig:9}
\end{figure}

 \vspace{-0.2cm}
The   ARBX(1) modelling is adopted since   pollution particles are mainly procured from natural sources (i.e., influenced by meteorological variables), or due to human activity. In our study we incorporate the exogenous information coming from meteorological variables. Specifically, we consider the following four exogenous variables ($b=4$ in our ARBX(1) model): \textcolor{blue}{d}aily average temperature ($^\circ C$), daily average atmospheric pressure ($hPa$), daily average wind speed ($ms^{-1}$), and daily maximum gradient of temperature ($^\circ C$), computed all of them from hourly measurements.
Note that daily maximum gradient of temperature denotes the daily maximum of the hourly differences between the temperature at $2$ and $100$ meters. \textcolor{red}{Wind speed is commonly included jointly with wind direction. However, this aspect would require an spatial correlation structure for the stations, which is out of the scope of the current article. Some discussion about how the spatial extension can be implemented can be found in Section \ref{sec:6}.}

\begin{figure}[H]
\centering
 \hspace{-0.9cm} \includegraphics[width=0.55\textwidth]{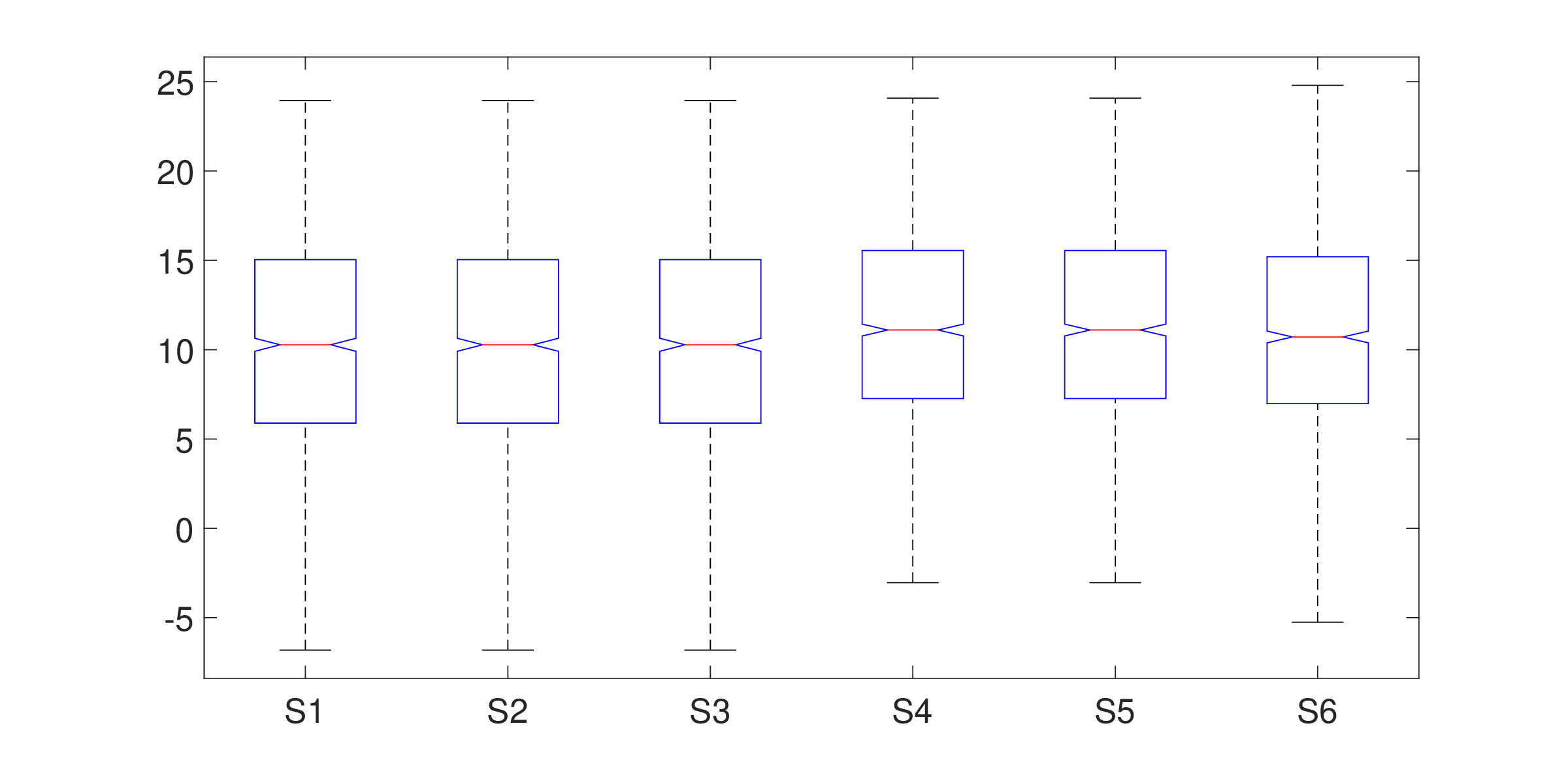}
 \hspace{-0.8cm} \includegraphics[width=0.55\textwidth]{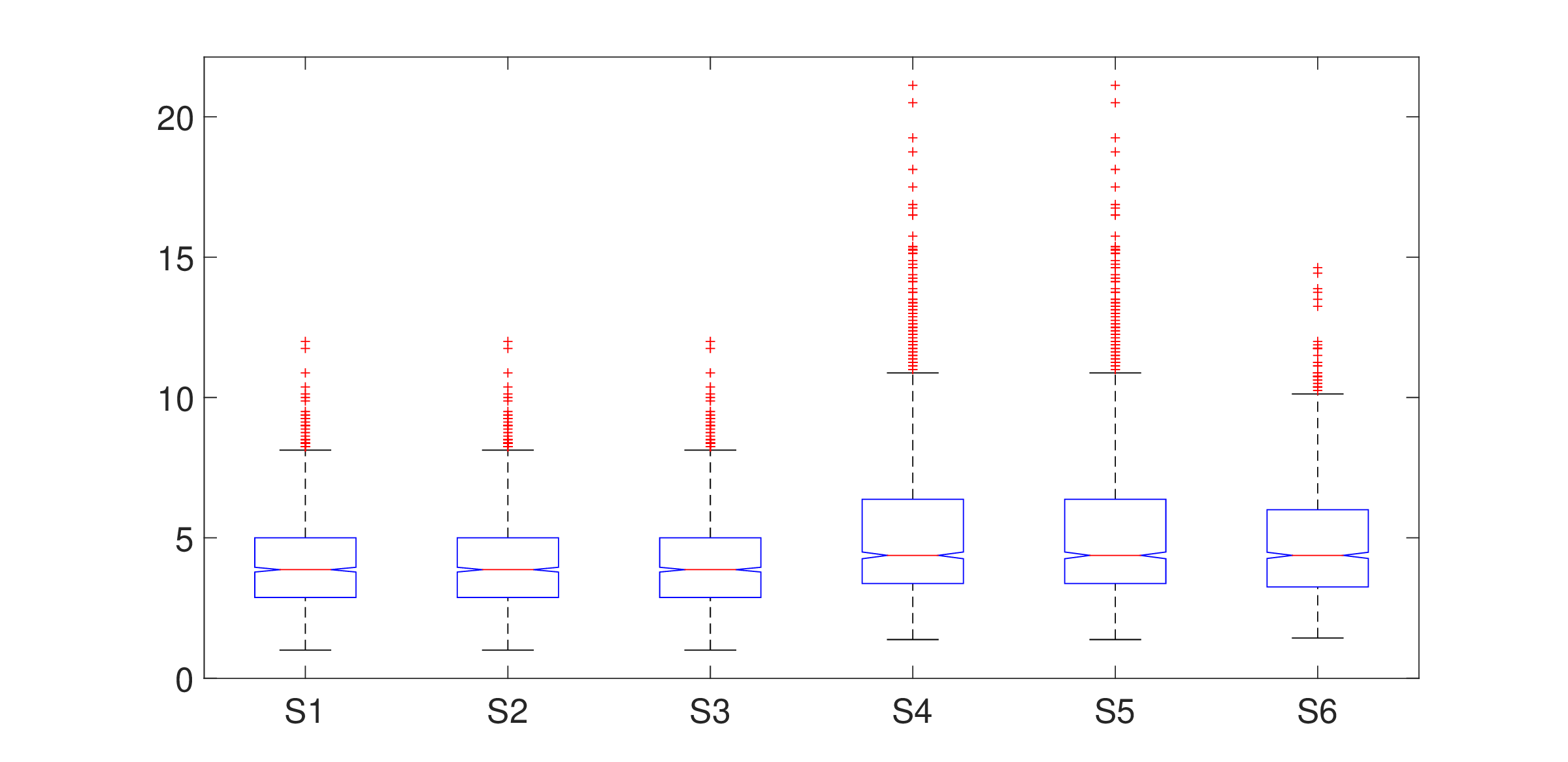}
  
 \hspace{-0.9cm}  \includegraphics[width=0.55\textwidth]{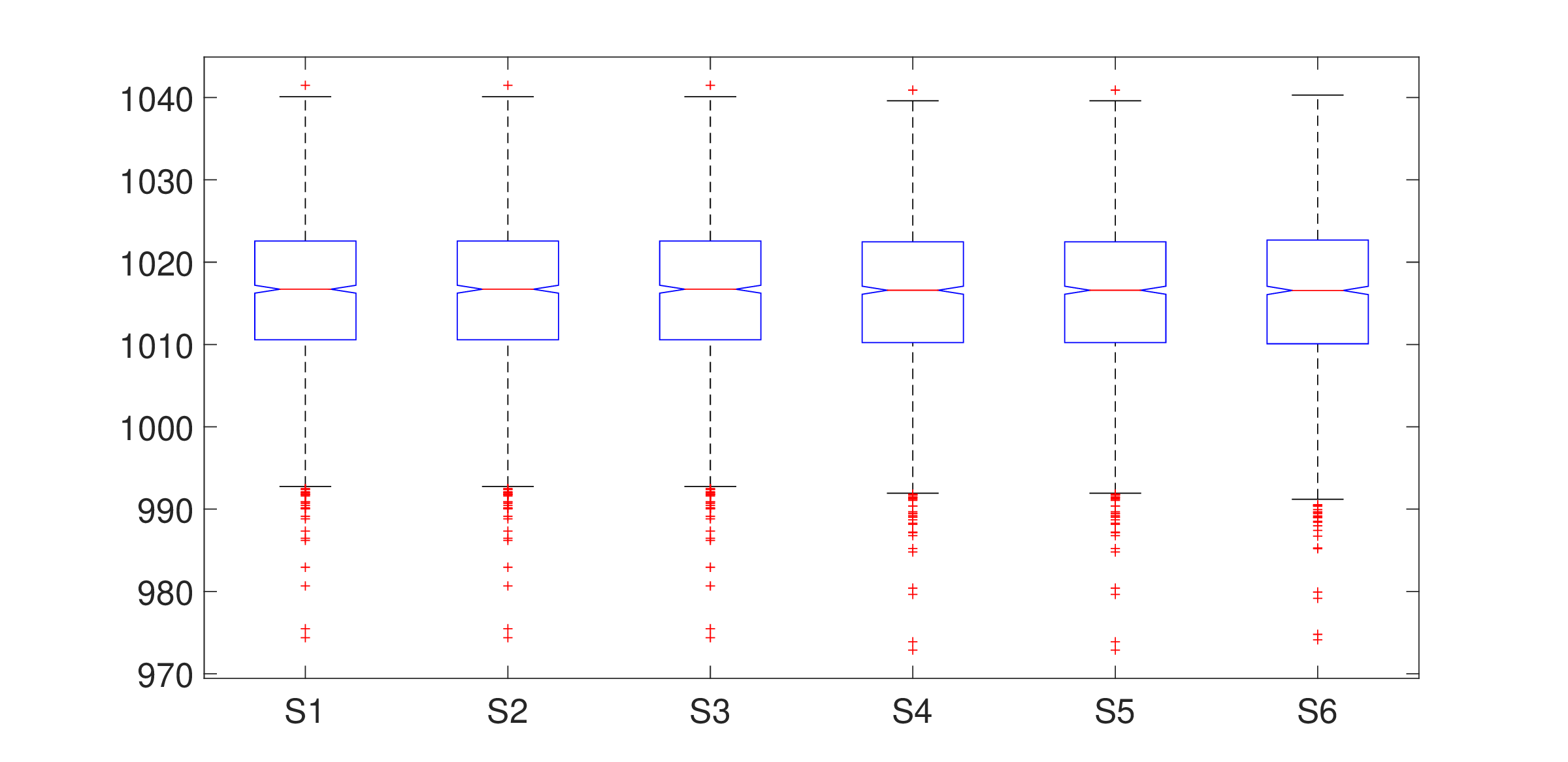} 
 \hspace{-0.8cm} \includegraphics[width=0.55\textwidth]{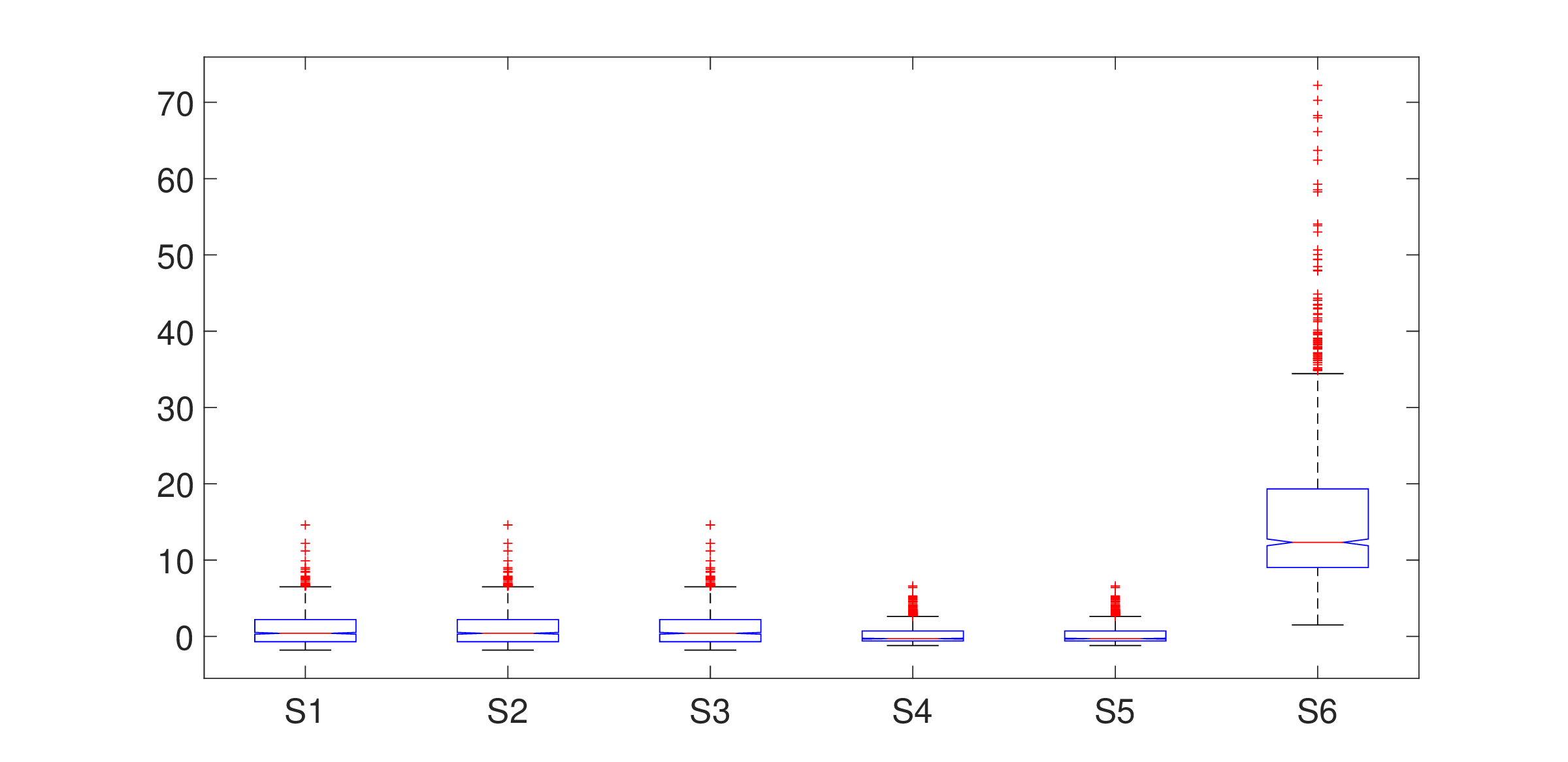}
  
   \vspace{-0.2cm}
 \caption{{\small \textcolor{red}{Boxplots reflecting the sample  behaviour of exogenous variables at each meteorological station (mean temperature on the top left, mean wind speed on the top right, mean air pressure at the bottom left and maximum temperature gradient at the bottom right), associated with  pollution stations $\left(S_1,~S_2,~S_3\right)$, $\left(S_4,~S_5\right)$  and $\left(S_6\right)$.}}}
 \label{fig:10}
\end{figure}

Measurements of these  meteorological variables were collected at three  meteorological stations belonging to the French national meteorological service, such that each air pollution station   is associated with the closest meteorological station. Thus, pollution stations 
$S_{1}, S_{2}$ and $S_{3}$ are associated to a common meteorological station.  Also, a second meteorological station covers the pollution stations  $S_{4}$ and $S_{5}.$ Finally, a third meteorological station is associated with  the pollution station $S_{6}.$ \textcolor{red}{Figure \ref{fig:10} displays boxplots for all of these measurements collected in the three meteorological stations available (see also Table \ref{tab:4} in the Appendix, displaying basic
  descriptive statistics about them).} As commented,  pollution  monitoring stations are separately analysed, and no spatial interaction is contemplated.

\subsection{Data preprocessing}
\label{sec:52}

\begin{figure}[H]
\centering
  \includegraphics[width=0.82\textwidth]{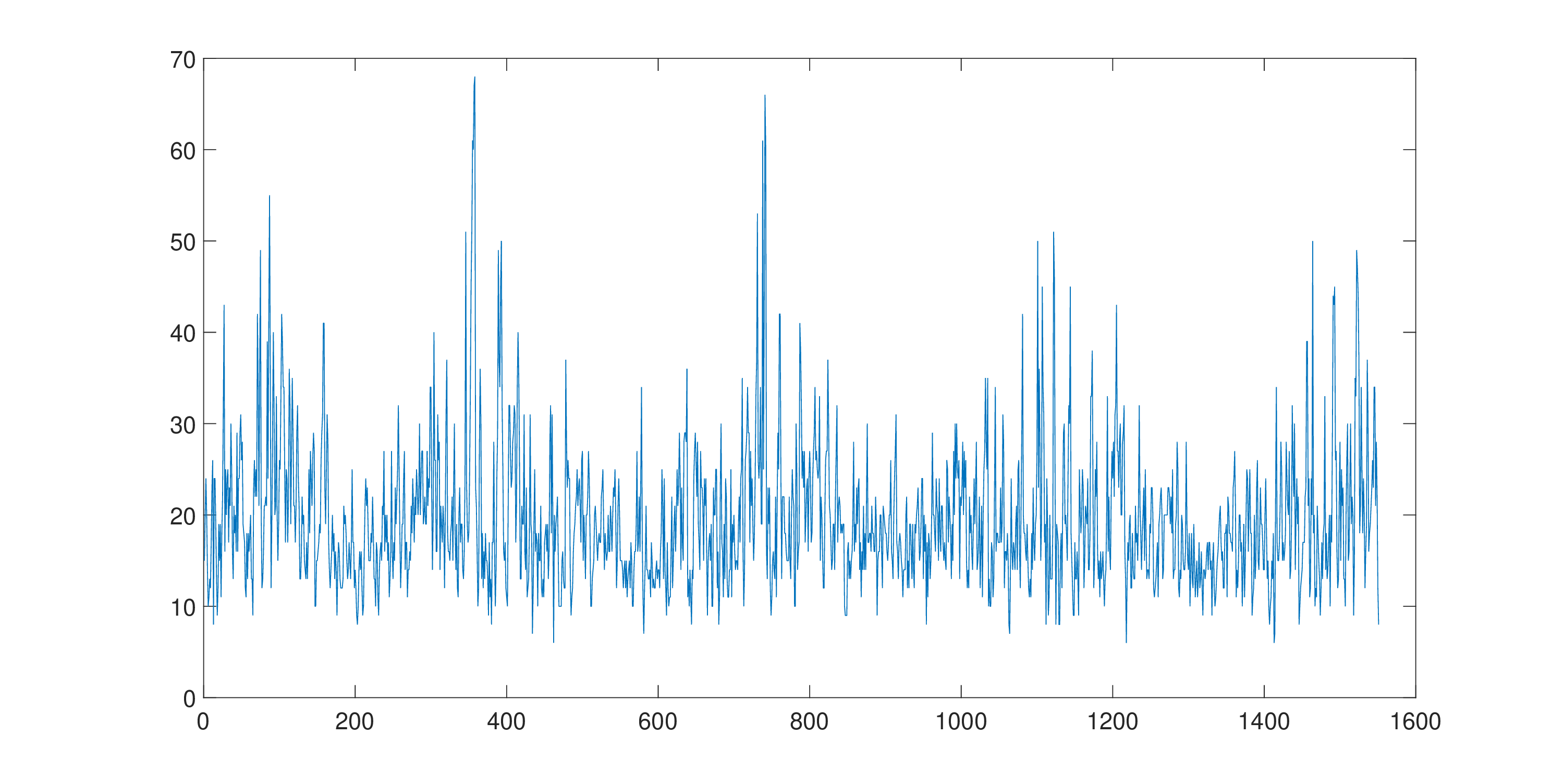}
  \vspace{-0.25cm}
 \caption{{\small \textcolor{blue}{Daily observations (from 01/01/2007 to 31/03/2011)  at  station $S_{1}$ after imputation.  }}}
 \label{fig:11hh}
\end{figure}

To obtain our functional dataset the following steps are implemented:

\vspace{0.3cm}
\noindent \textbf{Step 1: Missing--data imputation.}  \textcolor{red}{As can be checked in Table \ref{tab:3}  below,  missing values appeared: an imputation procedure is required. Taking the more powerful \textsc{R} package as a reference (see \cite{Moritz17,MoritzBartz17}), we have implemented one of the imputation methods there described, assigning to missing values an average of previous and posterior non missing values. This choice has been adopted for preserving the dependence structure of data. 
Since pollution data is strongly linked with routines and consumption patterns in business days, the past and the next five values are considered. At each station, $1551$ records are then available. Daily observations of the endogenous variable at station $S_{1}$ are displayed after imputation in  Figure  \ref{fig:11hh}.}

\vspace{-0.45cm}

\vspace{0.2cm}
\noindent \textbf{Step 2: Fitting to $\boldsymbol{31}$--days standard months.}  
 \textcolor{red}{As can be appreciated in the previous sections, our functional data should be evaluated in the same function spaces; that is, exogenous and endogenous functional variables should be valued in the same Banach space. In this way, aimed at achieving   balanced data for each of the stations, we need that all months (i.e., all intervals) contain the same amount  of days (i.e., the same amount of grid points). For this purpose, \textit{Cubicspline} option in  \texttt{fit.m}  \texttt{MatLab} function  has been applied, obtaining $31$ measurements per month: an unique grid of  $31 \times  51 = 1581$ points is considered for each of exogenous and endogenous variables, at each station.}

\vspace{0.3cm}

\noindent \textbf{Step 3: Splitting our dataset.} At each pollution station $S_{c},$ $c=1,\dots,6,$ we will construct our plug--in predictor $\widehat{\overline{X}}_{50}^{c}=\widetilde{\overline{\rho}}_{k_{n}}
(\overline{X}_{49}^{c}),$ from a functional sample $\{\overline{X}_{0},\dots,\overline{X}_{49}\}$ of size $50.$ Thus, functional prediction is achieved for the last month, March, 2011, at each pollution station.

\vspace{0.3cm}
\noindent \textbf{Step 4: Detrending and deseasonalizing.} \textcolor{red}{As usual, a polynomial trend $a_0 + a_1 t + a_2 t^2 +  a_3 t^3 + ...$ will be fitted for detrending all the curve data. Thus, we have checked the trends fitting  polynomials of increasing degree, stopping when the fitting trends between two successive degrees display a similar behaviour. As shown in Figure \ref{fig:trends} in the Appendix, and applying the law of parsimony, a polynomial quadratic trend $a_0 + a_1 t + a_2 t^2$  has finally been fitted (based on a functional sample of size $50$) for detrending all the curve  data, including the last month.  After detrending, annual seasonality is also removed}.

 \vspace{0.3cm}
\noindent \textbf{Step 5: Modelling by an  ARBX(1) process.} Summarizing, from the previous steps, our functional sample is constituted by $50$ detrended, and annually deseasonalized observed  curves for the endogenous and exogenous variables (on  a grid of $1581$ points).  Plug--in functional prediction is achieved for the $51$--th month, from the  observed curves at the previous $50$ months by fitting ARBX(1) model in equation (\ref{eq_74}) below. 
  Figure \ref{fig:12hh}  displays our PM$_{10}$ functional dataset at station $S_{1},$ corresponding to the period  
  January 2007--February 2011.

\begin{figure}[H]
\centering
 \includegraphics[width=0.99\textwidth]{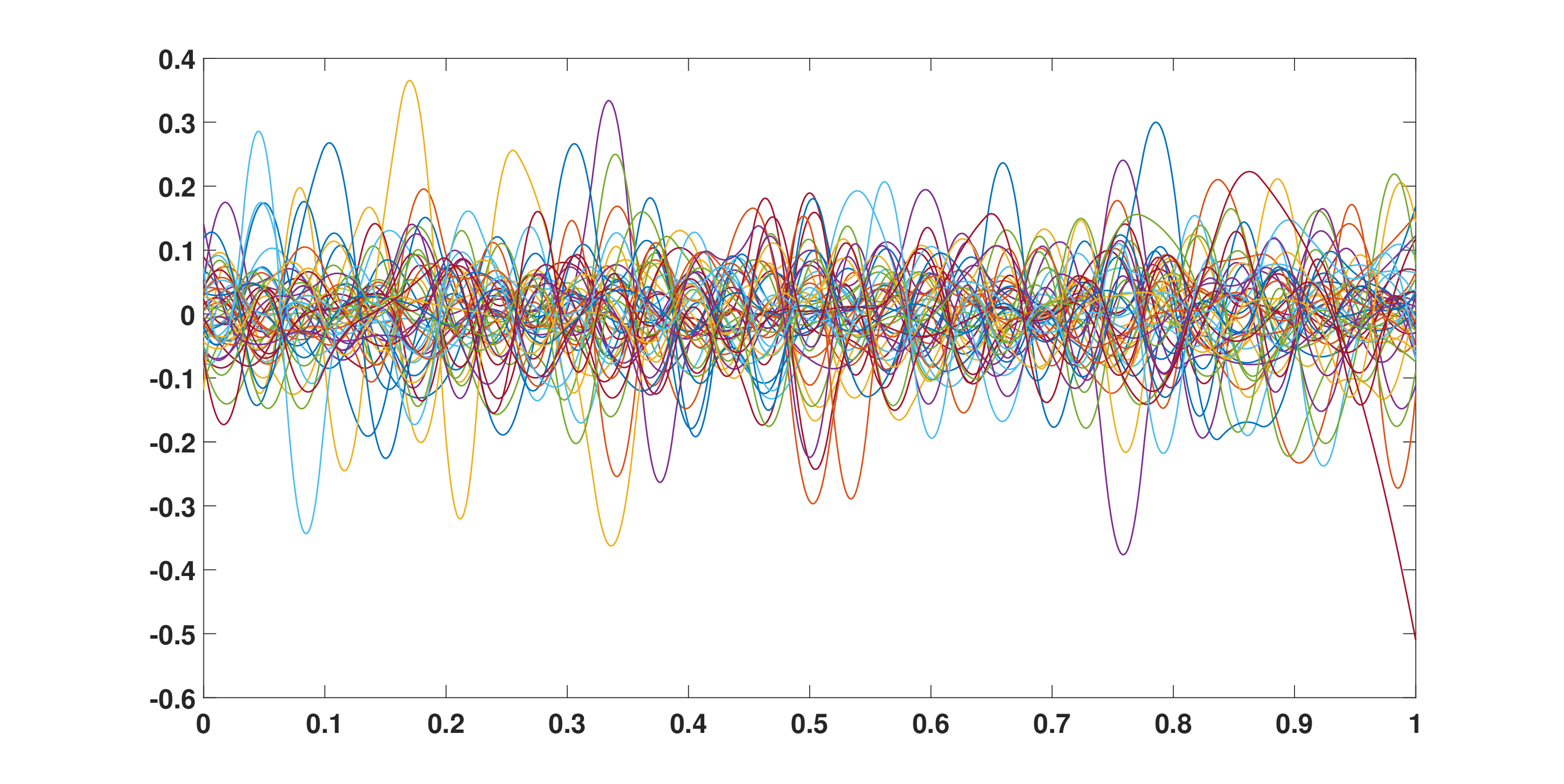}

\vspace{-0.2cm}
\caption{{\small Observed PM$_{10}$ curves  at stations $S_{1}$ for the period January 2007--February 2011.}}
 \label{fig:12hh}
\end{figure}

Specifically, the following ARBX(1) model is fitted:
\begin{equation}\overline{X}_{n}^{c}  = \overline{\rho}^{c} \left( \overline{X}_{n-1}^{c} \right) + \overline{\varepsilon}_{n}^{c}, \quad n=0,1,\ldots,T=49,\quad c=1,\ldots,6,\label{eq_74a}
\end{equation}
\noindent where  $b=4,$ and hence, 
\begin{equation}
\overline{X}_{n}^{c} = \begin{pmatrix} X_{n}^{c} \\ Z_{n+1,1}^{c} \\ Z_{n+1,2}^{c}  \\ Z_{n+1,3}^{c}  \\ Z_{n+1,4}^{c} \end{pmatrix}, \quad \overline{\varepsilon}_{n}^{c}= \begin{pmatrix} \varepsilon_{n}^{c} \\ \eta_{n,1}^{c} \\ \eta_{n,2}^{c} \\ \eta_{n,3}^{c} \\ \eta_{n,4}^{c} \end{pmatrix}, \quad 
\overline{\rho}^{c} = \begin{pmatrix} \rho^{c} & a_{1}^{c} & a_{2}^{c} &  a_{3}^{c} & a_{4}^{c} \\ \mathbf{0}_B & u_{1}^{c} & \mathbf{0}_{B}   &  \mathbf{0}_{B} &  \mathbf{0}_{B}  \\   \mathbf{0}_B & \mathbf{0}_{B}  & u_{2}^{c}  & \mathbf{0}_{B}  & \mathbf{0}_{B}  \\ \mathbf{0}_B & \mathbf{0}_{B}  & \mathbf{0}_{B}   & u_{3}^{c}  & \mathbf{0}_{B} \\ \mathbf{0}_B & \mathbf{0}_{B}  & \mathbf{0}_{B}   & \mathbf{0}_{B}   & u_{4}^{c} \end{pmatrix}, \quad c=1,\ldots,6.\label{eq_74}
\end{equation}

Equivalently, for   $c=1,\ldots,6,$ 
\begin{equation}
X_{n}^{c}  = \rho^{c} \left( X_{n-1}^{c} \right)  + \displaystyle \sum_{i=1}^{4} a_{i}^{c} \left(Z_{n,i}^{c}\right)  + 
\varepsilon_{n}^{c}, \quad  n=0,1,\ldots,T=49, \label{eq_90_}
\end{equation}
\noindent with, for $i=1,\dots,4,$   $a_{i}^{c} \in \mathcal{L}(B),$ $\rho^{c}\in \mathcal{L}(B),$ and
\begin{equation}
Z_{n,i}^{c}= u_{i}^{c} \left( Z_{n-1,i}^{c} \right) + \eta_{n,i}^{c}, \quad u_{i}^{c} \in \mathcal{L}(B), \quad  n=0,1,\ldots,T=49. 
\end{equation}

It can be observed in Figure \ref{fig:12hh} that 
PM$_{10}$ curves  are continuous. Hence, $\beta >1/2.$ \textcolor{red}{As before,} we look for the minimal local  regularity order, in our  fitting of the previous introduced ARBX(1) model, with $b=4.$ Thus,  the parameter values $\beta = 3/5$ and $\gamma= \gamma_{i} = 2\beta + \epsilon,$ with $\epsilon = 0.01,$ $i=1,2,3,4,$ have been considered in the definition of our function space scenario, given by  

\begin{equation}
\overline{B}=\left[\mathcal{B}_{\infty,\infty}^{0} \left([0,1]\right)\right]^{5}, \ \widetilde{\overline{H}}=\left[\mathcal{B}_{2,2}^{-\beta} \left([0,1]\right)\right]^{5}, \ \overline{H}=[L^{2}([0,1])]^{5},\ \mathcal{H}(\overline{X})=[\mathcal{B}^{\gamma}_{2,2}\left([0,1]\right)]^{5}.
\label{banachcontext}
\end{equation}

\subsection{The performance of the ARBX(1)  plug--in predictor}
\label{sec:44}

In the implementation of the leave--one--out cross validation procedure, at each pollution station $S_{c},$ $c=1,\dots,6,$  the following functional sample is considered:
\begin{eqnarray}
\overline{Y}^{h,c} &=& \left\lbrace \overline{Y}_{i}^{h,c}, \quad i=0,1,\ldots,47 \right\rbrace = \left\lbrace \overline{X}_{i}^{c},  \quad i=0,1,\ldots,48 \right\rbrace \setminus \left\lbrace \overline{X}_{h}^{c} \right\rbrace, \nonumber
\end{eqnarray}
\noindent by \textcolor{blue}{removing} the functional data $\overline{X}_{h}^{c},$  $h=0,1,\ldots,48,$ as well as the functional data $\overline{X}_{49}^{c},$ $c=1,\dots,6,$ in the computation of the componentwise estimator (\ref{estimator}) of the autocorrelation operator $\overline{\rho}.$ Thus, at each iteration $h\in \{0,1,\ldots,48\}$ of the implemented leave--one--out cross validation procedure, the ARBX(1) 
plug--in   predictor  $[\widehat{\overline{X}}_{50}^{c}]_{h,k_n}$   of $\overline{X}_{50}^{c}$ is computed, from a functional sample of size $48,$ considering  the truncation order $k_{n},$ as follows
\begin{eqnarray}
 [\widehat{\overline{X}}_{50}^{c}]_{h,k_n}  &=& \displaystyle \sum_{j,l=1}^{k_n} \frac{1}{\widetilde{C}_{n,j}^{c,h}} \langle \overline{X}_{49}^{c,h}, \widetilde{\phi}_{n,j}^{c,h} \rangle_{\widetilde{\overline{H}}  } \left(\frac{1}{47} \displaystyle \sum_{i=0}^{46} \langle \overline{Y}_{i+1}^{c,h}, \widetilde{\phi}_{n,j}^{c,h} \rangle_{\widetilde{\overline{H}}} \langle \overline{Y}_{i}^{c,h}, \widetilde{\phi}_{n,l}^{c,h} \rangle_{\widetilde{\overline{H}}} \widetilde{\phi}_{n,l}^{c,h} \right), \nonumber\\
 \label{pp50}
\end{eqnarray}
\noindent  for   $c=1,\ldots,6.$ Here, the empirical eigenvalues   $$\left\lbrace \widetilde{C}_{n,j}^{c,h},\ j=1,\dots,48,\  ~h=0,1,\ldots,48,\ c= 1,\ldots,6 \right\rbrace$$ \noindent are  calculated by the formula
\begin{equation}
\widetilde{C}_{n,j}^{c,h} = \frac{1}{48} \displaystyle \sum_{i=0}^{47} \langle \overline{Y}_{i}^{c,h}, \widetilde{\phi}_{n,j}^{c,h} \rangle_{\widetilde{\overline{H}}}^{2} = \frac{1}{48} \displaystyle \sum_{\substack{i \neq h \\ i=0}}^{48} \langle \overline{X}_{i}^{c}, \widetilde{\phi}_{n,j}^{c,h} \rangle_{\widetilde{\overline{H}}}^{2}. \nonumber
\end{equation}

\begin{figure}[H]
\centering
 \hspace{-0.9cm} \includegraphics[width=0.55\textwidth]{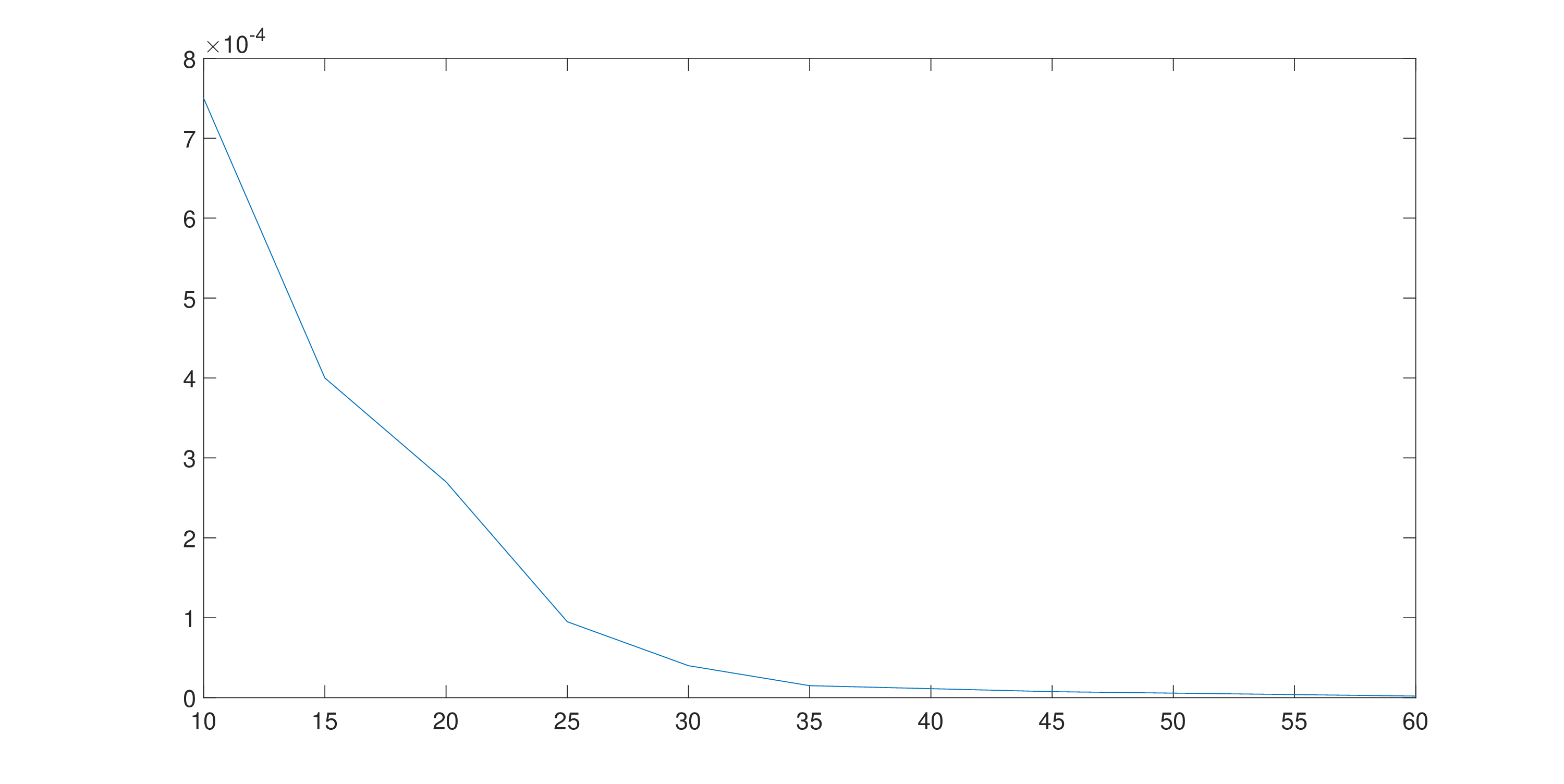}
\hspace{-0.7cm}\includegraphics[width=0.55\textwidth]{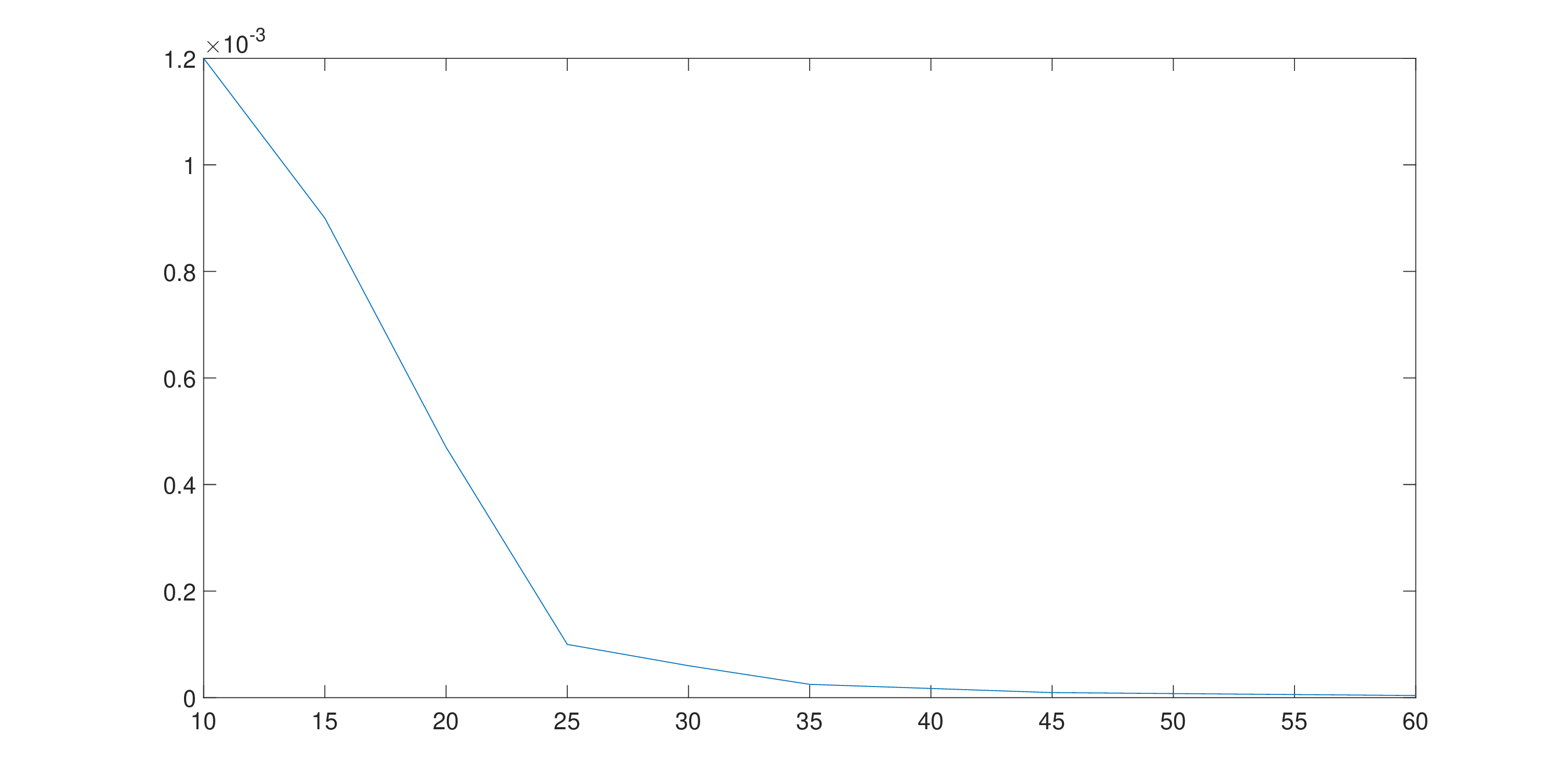} 

 \vspace{-0.2cm}
 \caption{{\small Evaluation at stations $S_1$ (top) and $S_6$ (bottom) of the empirical norm   $L_k = \sup_{\overline{x}\in \overline{B},\ \|\overline{x}\|_{\overline{B}}\leq 1}\left\| \overline{\rho} \left( \overline{x} \right) - \displaystyle \sum_{j=1}^{k} \langle \overline{\rho} \left(\overline{x} \right), \overline{\phi}_{n,j} \rangle_{\widetilde{\overline{H}}} \overline{\phi}_{n,j} \right\|_{\overline{B}}$, for values $k=10,15,20,25,30,35,45,60,$ displayed in the horizontal axis. }}
 \label{fig:12}
\end{figure}

\noindent In the above equations, for each $c=1,\ldots,6$, and $h=0,1,\ldots,48,$ $\left\lbrace \widetilde{\phi}_{n,j}^{c,h},\ j\geq 1  \right\rbrace$ denotes the system of eigenvectors of the extended empirical autocovariance operator, based on a functional sample of size $48.$
The truncation parameter values  $k_{n,1} =  \left[ \log_2\left(\sqrt{n} \right) \right]^{-}$ and $k_{n,2} =  \left[ \ln\left( n^{5/2} \right) \right]^{-}$ are tested, in the computation of (\ref{pp50}).  Similarly to the simulation study undertaken, all the required conditions for the strong--\textcolor{blue}{consistency} are checked. 
 In particular,  \textbf{Assumptions A4--A5}  directly follow  from the function space scenario (\ref{banachcontext}) adopted. \textcolor{blue}{In addition}, Figure \ref{fig:12} displays the convergence to zero required in \textbf{Assumption A3}, for the pollution stations $S_1$ and $S_6.$

Figure  \ref{fig:errors} below displays  the mean leave--one out cross validation functional errors 
\begin{eqnarray}
E_{c}^{k_{n,m}} &=&  \frac{1}{49} \displaystyle \sum_{h=0}^{48} \left\| \overline{X}_{50}^{c} - \left[\widehat{\overline{X}}_{50}^{c} \right]_{h,k_{n,m}}  \right\|_{\overline{B}}, \quad c = 1,\ldots,6, \quad m=1,2, \label{error_real_data}
\end{eqnarray}
\noindent at the six  pollution stations studied, for the two truncation orders analysed. In the calculation of (\ref{error_real_data}),  the   Besov and Sobolev norms involved in our function space scenario  
(\ref{banachcontext})  are computed by projection into Daubechies wavelets of order $N=10$ (see \cite{Daubechies92}),  with six resolution levels (\textcolor{red}{see also  Figure \ref{fig:2} above}).

\begin{figure}[H]
\begin{center}
 \includegraphics[width=0.95\textwidth]{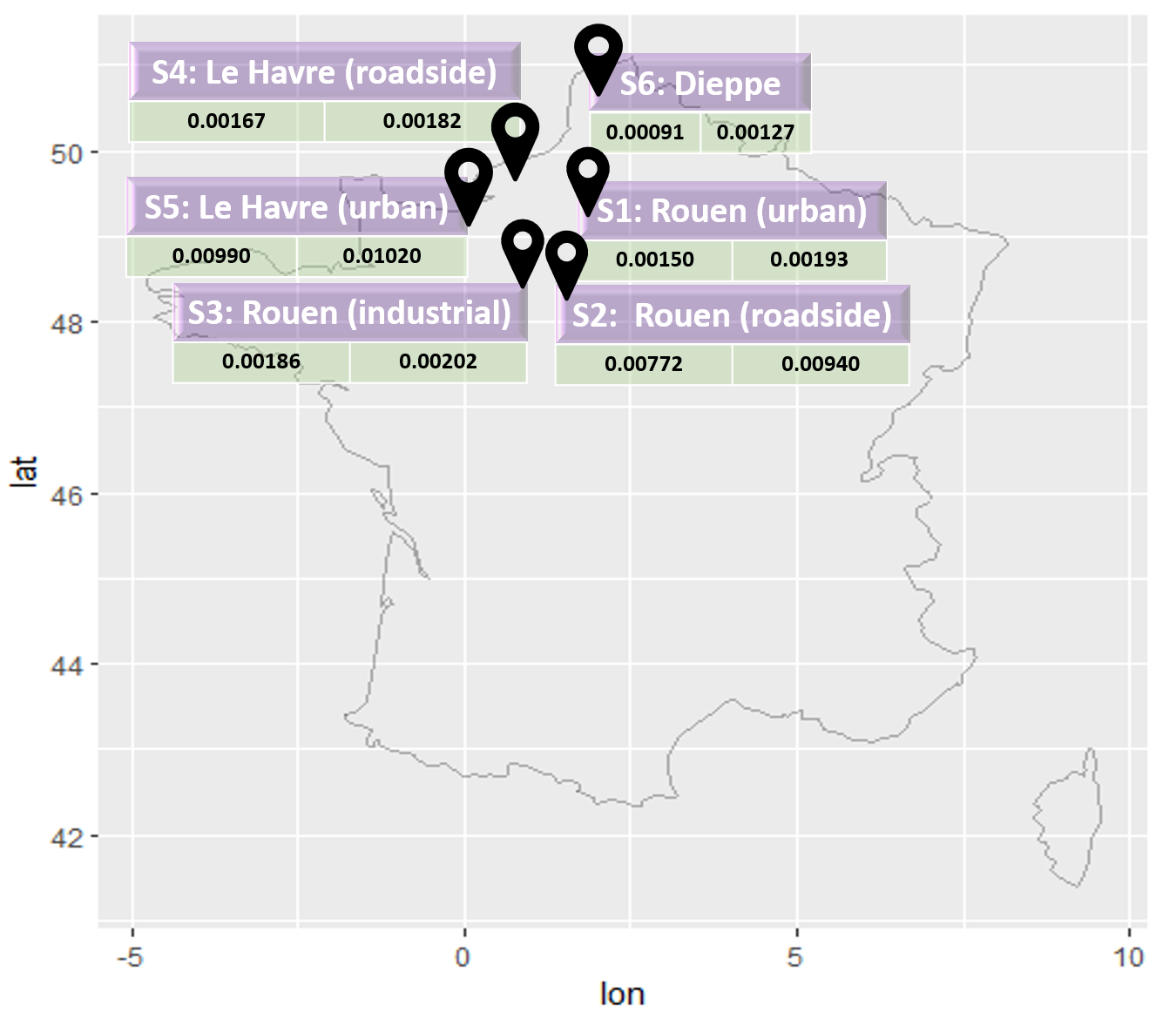}
  \vspace{-0.2cm}
 \caption{{\small  Map displaying mean leave--one out cross validation functional errors (\ref{error_real_data})  at pollution stations $\left\lbrace S_c, \ c=1,\ldots,6 \right\rbrace.$  The truncation parameters $k_{n,1} =  \left[ \log_2\left(\sqrt{n} \right) \right]^{-}$ (on left) and $k_{n,2} =  \left[ \ln\left( n^{5/2} \right) \right]^{-}$ (on right) have been tested.}}
 \label{fig:errors}
\end{center}
\end{figure}


In Figure  \ref{fig:errors}, the worst performance is observed at pollution  stations $S_2$ and
$S_5,$ corresponding to roadside stations. As commented before, the traffic flow is one of the main factors inducing  the higher--variability  displayed by PM$_{10}$ concentrations in these stations (\textcolor{red}{see Table \ref{tab:3} in the Appendix}).  A slightly better performance can be observed with truncation order $k_{n,1},$  but, indeed,  a significant improvement cannot be concluded. When  larger values of parameter $\beta ,$  and hence, of parameter $\gamma ,$ defining   our function space scenario, are considered,  a stronger smoothing of our original data set is achieved,  in terms of Sobolev  and Besov norms.  Thus, a better performance is obtained, with  a precision loss, in the approximation of the local  \textcolor{blue}{behaviour} of PM$_{10}$ concentrations.


\section{Final comments}
\label{sec:6}

It is well-known that FDA techniques provide a flexible framework for the local analysis of high--dimensional data which are continuous in nature. One of the main subjects in FDA is the suitable choice of the function space, where the observed data take their values. In particular, the norm of the selected space should  provide an accurate measure of the local \textcolor{blue}{variability}  of the  observed endogenous and exogenous   variables,  that could be crucial in the posterior representation of the possible  interactions  with   the phenomena of interest and its evolution. That is the case of the real-data example analysed \textcolor{red}{in Section \ref{sec:5}.} 

This paper adopts  an abstract Banach space framework, assuming an autoregressive dynamics in time, for all the functional random variables involved in the model.  Specifically, an ARBX(1) model is considered. The endogenous and exogenous information affecting the functional response at a given time is incorporated through a  suitable  linear model, involving a matrix autocorrelation operator. This operator models possible interactions between all endogenous and exogenous functional random variables at any time. 

Particularly, the scale of fractional Besov spaces provides a suitable functional framework, where the presented approach can be implemented, modelling local regularity/singularity in an accurate way. Note that the norms in these spaces can be characterised in terms of the wavelet transform. Specifically, wavelet bases provide countable dense systems in Besov spaces, that can be used in the definition of the inner product and associated norms in weighted fractional Sobolev spaces, constructed from the space of square integrable functions on an interval (see \cite{Triebel83}, and \cite{RuizAlvarez18_JMVA}).   Thus, suitable embeddings can be established for applying the construction in Lemma 2.1 in \cite{Kuelbs70}. 
As special cases of well-known Banach spaces within our framework,  we refer to $\mathcal{C}([0,1])$ the space of continuous functions on $[0,1],$ with the supremum norm, and $\mathcal{D}([0,1])$ the  Skorokhod space of right-continuous functions on the interval $[0,1],$ having
a left limit at all $t\in [0,1].$  Note that these spaces have been  widely used in the FDA literature in a Banach--valued time series context (see \cite{Bosq00}).

\medskip

The simulation study and real--data application illustrate the fact that our approach is sufficiently flexible to describing  the local \textcolor{blue}{behaviour} of both,  regular and   singular functional  data. Note that, in the singular case, we can choose a suitable norm that \textcolor{blue}{measures} the local fluctuations in a precise way. This information is relevant, for example, in the analysis of  PM$_{10}$ concentrations, as was illustrated  in \textcolor{red}{Section \ref{sec:5}}. An individual statistical analysis has been performed  in \textcolor{red}{Section \ref{sec:5}} at each pollution station. The incorporation of spatial interactions in the analysis could be addressed in a multivariate infinite-dimensional spatial  framework, and constitutes the subject of a subsequent paper.

\medskip

\textcolor{red}{Concerning extending this methodology to a spatiotemporal framework,...}

\section*{APPENDIX: FIGURES}

\begin{figure}[H]
\centering
\includegraphics[width=0.99\textwidth]{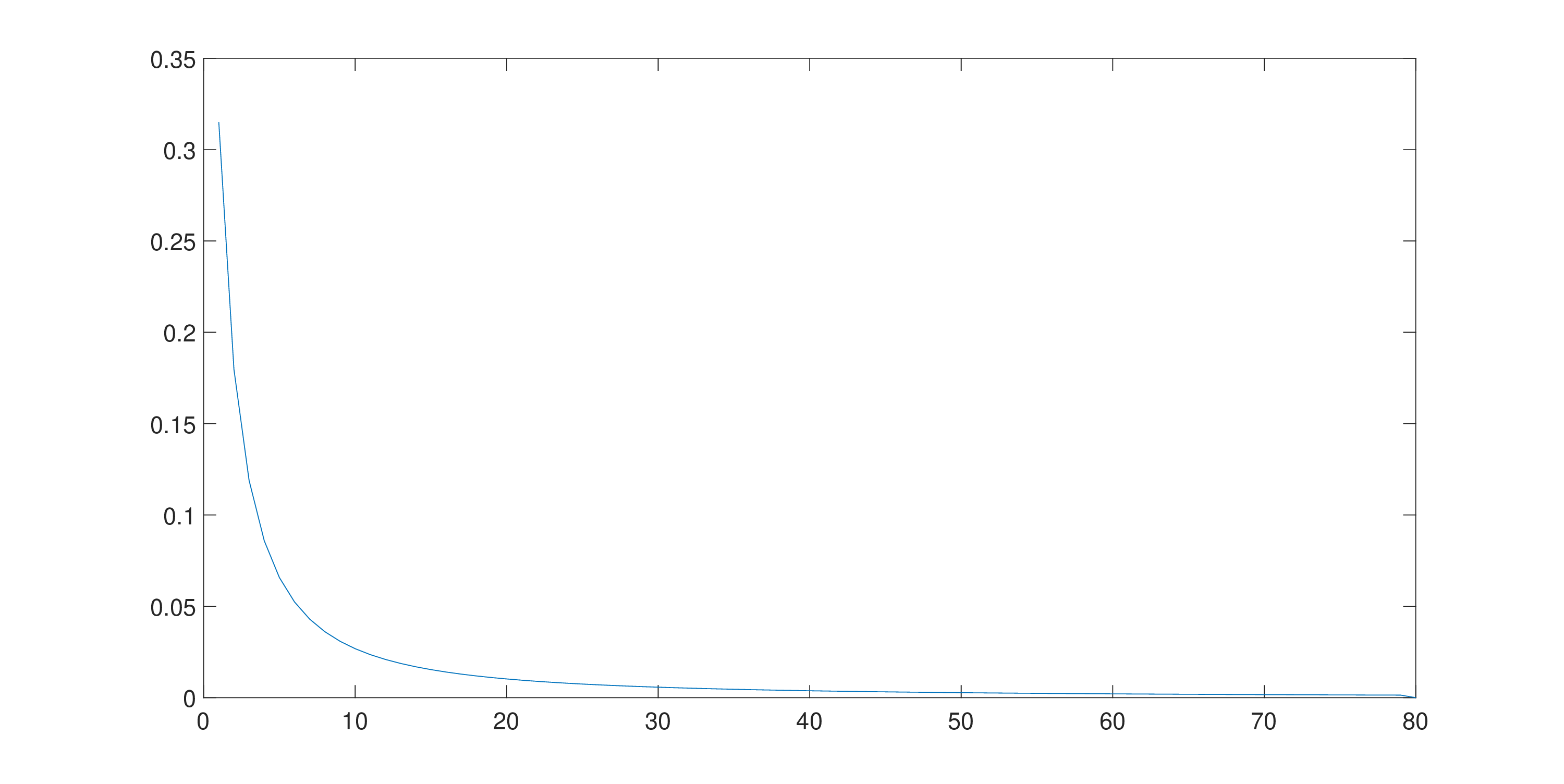}
  \vspace{-0.2cm}
 \caption{{\small Values of $\displaystyle \left\| \overline{\rho} \left( \cdot \right) - \displaystyle \sum_{l=1}^{K} \langle \overline{\rho} \left( \cdot \right), \overline{\phi}_j \rangle_{\left[ \mathcal{B}_{2,2}^{-\beta }([0,1])\right]^{b+1}} \overline{\phi}_j \right\|_{\mathcal{L}\left([\mathcal{B}_{\infty,\infty}([0,1])]^{b+1}\right)},$ for $K$--values  reflected in the horizontal axis.}}
 \label{fig:2h}
\end{figure}

 \begin{figure}[H]
\centering
 \includegraphics[width=0.99\textwidth]{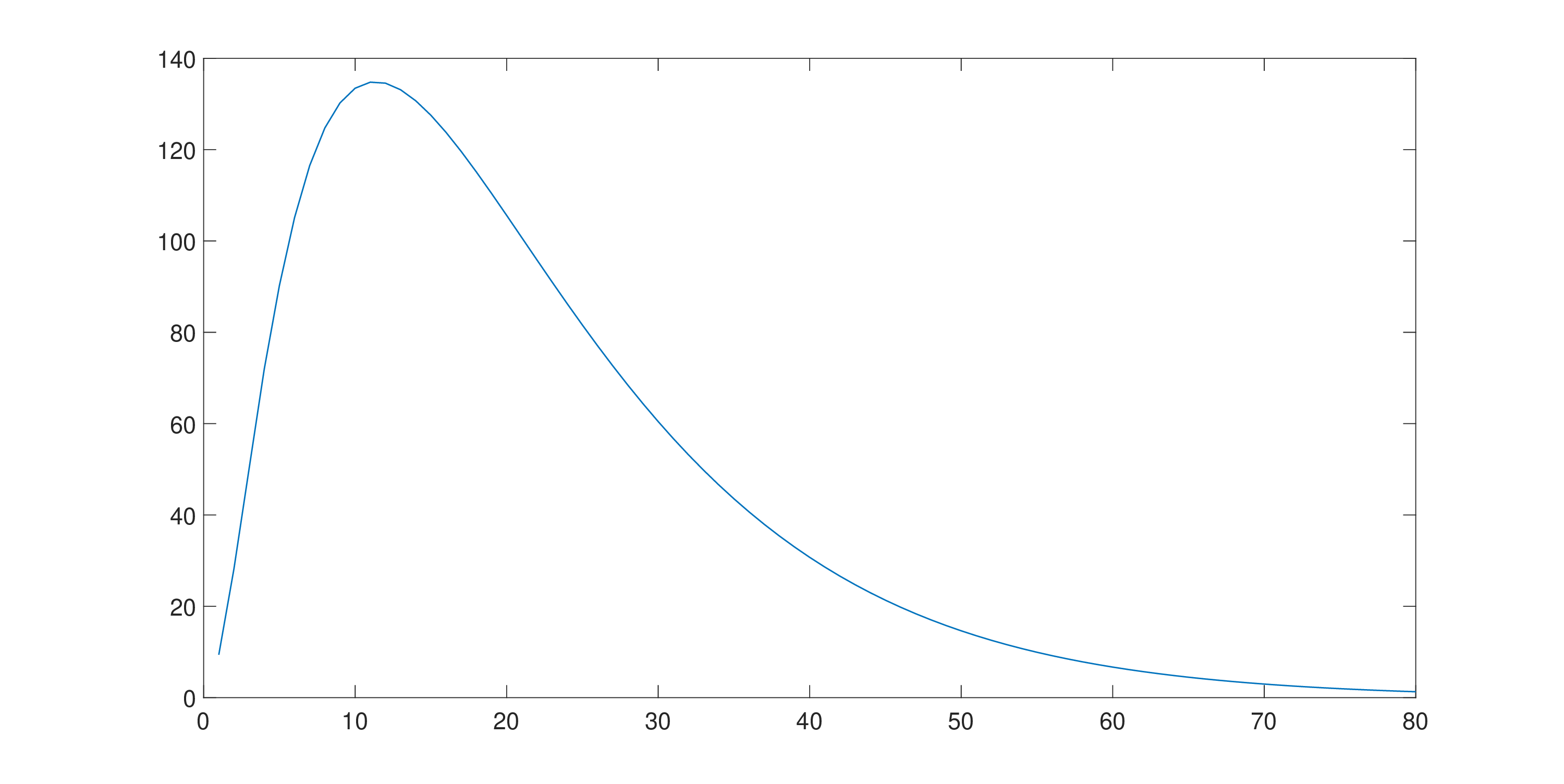}
 \vspace{-0.2cm}
 \caption{{\small Values of $\frac{k_n C_{k_{n}}^{-1} \displaystyle \sum_{j=1}^{k_{n}} a_j }{\sqrt{n/\ln(n)} },$ considering $k_{n} = \left[ \ln(n)  \right]^{-},$  for $k_{n}$-values  reflected in the horizontal axis. }}
 \label{fig:3}
\end{figure}

\begin{figure}[H]
\centering
 \includegraphics[width=0.99\textwidth]{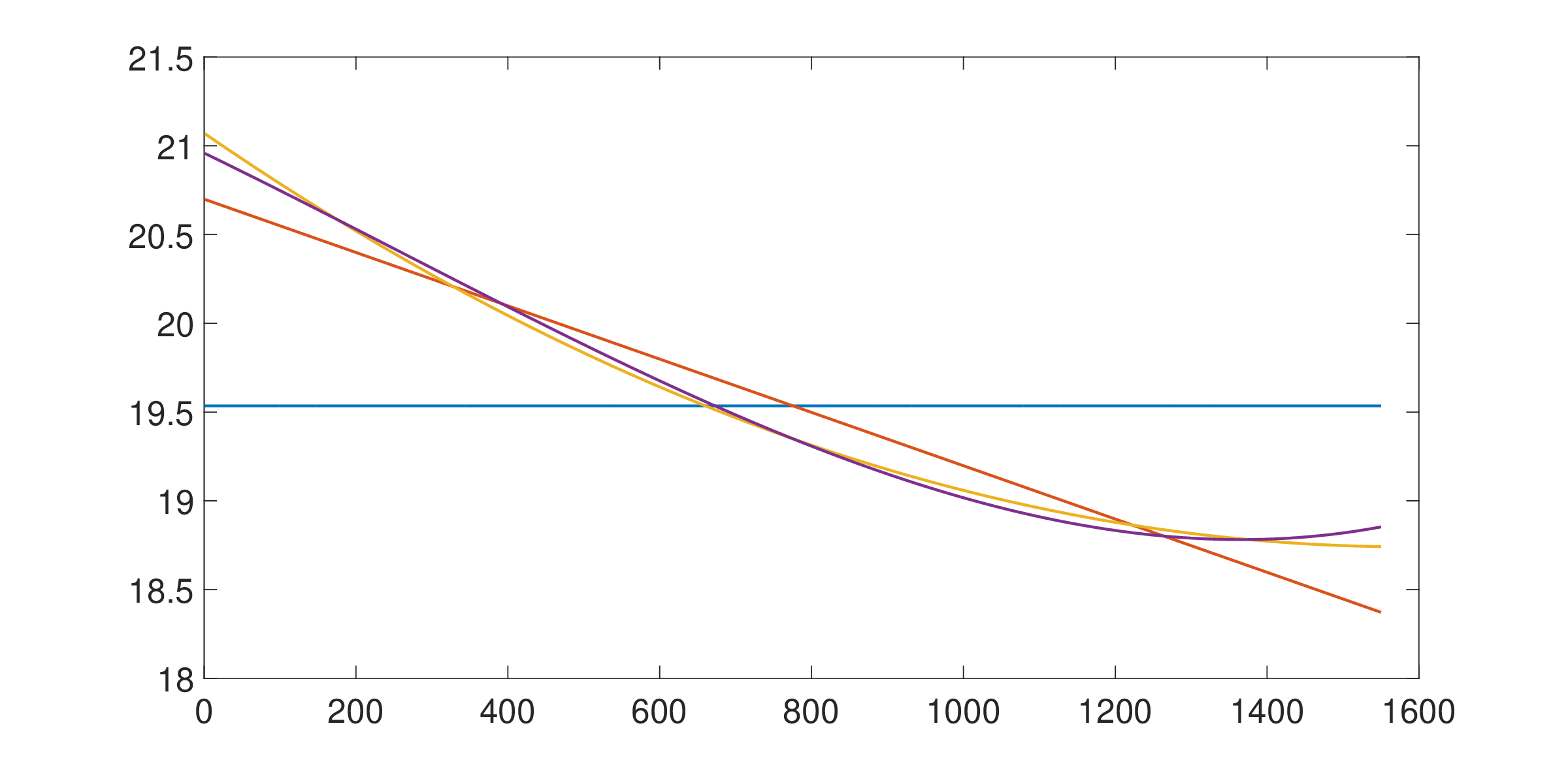}

\vspace{-0.2cm}
\caption{{\small \textcolor{red}{Fitting trends as polynomial of degrees $0$ (null trend; blue line), $1$ (linear trend; red line), $2$ (quadratic trend; yellow line) and  $3$ (cubic trend; purple line).}}}
 \label{fig:trends}
\end{figure}

\section*{APPENDIX: TABLES}

\begin{table}[H]
\caption{{\small Descriptive statistics from the    daily average PM$_{10}$ concentrations $\mu g~m^{-3}$ (the endogenous variable to be predicted) at the six pollution stations selected.}}
\centering
\begin{small}
\begin{tabular}{|c|c|c|c|c|c|c|c|}
  \hline
   $S_c$  & Location  & Min & Mean & Max & Standard Dev. & \%~missing \\
  \hline 
 $S_1$ & Rouen (Urban) &  6  & 19.6 & 68 & 7.9 & 1  \\
  \hline
 $S_2$ & Rouen (Roadside) & 7   & 24.6 & 90 & 9.3 & 0.4 \\
  \hline
  $S_3$ & Rouen (Industrial) & 4  & 18.9 & 80 & 8.8 & 1.4  \\
  \hline
    $S_4$ & Le Havre (Urban) & 5  & 19.2 & 86 & 8.5 & 3  \\
  \hline
     $S_5$ & Le Havre (Roadside) & 8   & 25.5 & 86 & 9.6 & 1.2  \\
  \hline
    $S_6$ & Dieppe (Rural) & 4  & 16.3 & 58& 6.3 & 1.7  \\
  \hline
\end{tabular}
  \label{tab:3}
\end{small}
\end{table}

\begin{table}[H]
\caption{{\small Descriptive statistics of daily average temperature ($T_m$), atmospheric pressure ($PA_m$),  wind speed ($VV_m$) and  maximum gradient of temperature ($GT_{max}$) in the three meteorological stations.}}
\centering
\begin{small}
\begin{tabular}{|c||c|c|c|c|}
  \hline 
    $\left(S_1,~S_2,~S_3\right)$ & $T_m$ & $VV_m$ & $PA_m$ & $GT_{max}$ \\
  \hline 
 Min  & -6.81& 1 &  974.36 & -1.8 \\
  \hline
 Median  & 10.28 & 3.8661 & 1016.71 & 0.4 \\
  \hline
  Mean  & 10.11 & 4.1349 & 1016.25 & 1.01 \\
  \hline
    Max  & 23.95 & 12 & 1041.49 & 14.6 \\
  \hline
     Standard Dev. & 6.05 & 1.65 & 9.74 & 2.17 \\
  \hline
\end{tabular}

\vspace{0.17cm}
\begin{tabular}{|c||c|c|c|c|}
  \hline
    $\left(S_4,~S_5 \right)$ & $T_m$ & $VV_m$ & $PA_m$ & $GT_{max}$ \\
  \hline   
 Min  & -3.04 & 1.38 & 972.88 & -1.2  \\
  \hline
 Median  & 11.1 & 4.38 & 1016.59 & -0.3 \\
  \hline
  Mean  & 11.03 & 5.26 & 1015.86 & 0.19 \\
  \hline
    Max  & 24.08 & 21.13 & 1040.9 & 6.6 \\
  \hline
     Standard Dev.   & 5.31 & 2.88 & 10.07 & 1.16 \\
  \hline
\end{tabular}

\vspace{0.17cm}
\begin{tabular}{|c||c|c|c|c|}
  \hline 
    $S_6$ & $T_m$ & $VV_m$ & $PA_m$ & $GT_{max}$ \\
  \hline 
 Min  & -5.25 & 1.43 & 974.13 & 1.49  \\
  \hline
 Median  & 10.71 & 4.38 & 1016.56 & 12.32 \\
  \hline
  Mean  & 10.64 & 4.85 & 1015.87 & 15.44 \\
  \hline
    Max  & 24.8 & 14.63 & 1040.3 & 72.24 \\
  \hline
     Standard Dev.   & 5.36 & 2.13 & 10.05 & 9.86 \\
  \hline
\end{tabular}
  \label{tab:4}
\end{small}
\end{table}

\bigskip
  

\begin{acknowledgements}
 This work was supported in part by project MTM2015--71839--P (co-funded by Feder funds), of the DGI, MINECO, Spain.

\end{acknowledgements}




\bibliographystyle{spmpsci}      


\end{document}